\documentclass[prd,showkeys,floatfix,twocolumn,amsmath,amssymb,floatfix]{revtex4}
\usepackage{amsmath}
\usepackage{graphicx}
\usepackage{float}
\usepackage{epstopdf}
\usepackage{multirow}
\usepackage{subfigure}
\usepackage{color}
\usepackage{dashrule}
\usepackage[colorlinks,citecolor=blue,anchorcolor=red,menucolor=red,linkcolor=red,filecolor=red,runcolor=red,urlcolor=blue,frenchlinks=red]{hyperref}
\usepackage{orcidlink}
\usepackage{slashed}

\allowdisplaybreaks[3]

\newcounter{cnt}
\makeatletter

\let\oldhypertarget\hypertarget
\renewcommand{\hypertarget}[2]{%
  \oldhypertarget{#1}{#2}%
    \protected@write\@mainaux{}{%
        \string\expandafter\string\gdef
          \string\csname\string\detokenize{#1}\string\endcsname{#2}%
    }%
  }
\newcommand{\myhyperlink}[1]{%
  \hyperlink{#1}{\csname #1\endcsname}%
  }
\makeatother

\newcommand{\cref}[1]{\myhyperlink{#1}}

\newcommand{\etal}{\textit{et al}. }
\renewcommand{\arraystretch}{1.8}

\begin{document}


\title{Low-lying doubly charmed strange baryons via QCD sum rules}

\author{Hui-Min Yang$^1$}
\email{hmyang@pku.edu.cn}
\author{Hua-Xing Chen$^2$}
\email{hxchen@seu.edu.cn}
\author{Xuan Luo$^2$}
\email{cnxluo@seu.edu.cn}

\affiliation{
	$^1$School of Physics, Henan Normal University, Henan 453007, China \\
	$^2$School of Physics, Southeast University, Nanjing 210094, China
}

\begin{abstract}
	we systematically investigate the masses of the low-lying doubly heavy baryons and their $B^*(3/2^+)\to B(1/2^+)\gamma$ radiative transitions within QCD sum rules and light-cone sum rules (LCSR). By constructing mixed interpolating currents, we obtain $M_{\Xi_{cc}}=3.62^{+0.14}_{-0.15}$ GeV, $M_{\Xi_{cc}^*}=3.72^{+0.14}_{-0.13}$ GeV, $M_{\Omega_{cc}}=3.72^{+0.13}_{-0.15}$ GeV, and $M_{\Omega_{cc}^*}=3.82^{+0.13}_{-0.15}$ GeV, together with their doubly bottom partners around $10.27$--$10.34$ GeV. The extracted masses are compatible with the observed $\Xi_{cc}$ and $\Omega_{cc}$ states and predict hyperfine splittings of about $101^{+27}_{-11}$ MeV and $99^{+11}_{-12}$ MeV for the $\Xi_{cc}$ and $\Omega_{cc}$ systems, respectively. Using these inputs, we derive the LCSR for the electromagnetic form factors and obtain radiative widths $\Gamma(\Xi_{cc}^{*++}\to\Xi_{cc}^{++}\gamma)=5.25^{+8.57}_{-3.62}$ keV, $\Gamma(\Xi_{cc}^{*+}\to\Xi_{cc}^{+}\gamma)=0.31^{+0.84}_{-0.25}$ keV, and $\Gamma(\Omega_{cc}^{*+}\to\Omega_{cc}^{+}\gamma)=0.41^{+0.67}_{-0.30}$ keV, while the corresponding doubly bottom widths are at the $0.002$--$0.01$ keV level. We find that the magnetic-dipole contribution dominates, whereas the electric-quadrupole form factor is numerically small. The light- and heavy-quark photon-emission terms can interfere or enhance significantly, and the heavy-quark contribution cannot be neglected in radiative transitions. These results provide useful theoretical guidance for future searches for spin-$3/2$ doubly heavy baryons through radiative decay modes at LHCb, Belle II, and other experimental facilities.
\end{abstract}


\maketitle

\section{Introduction}
Doubly heavy baryons, composed of two heavy quarks and one light quark, provide an ideal platform  for studying the interaction between heavy quark symmetry and the nonperturbative dynamics of the light degrees of freedom. In particular, the two heavy quarks form a color-antitriplet subsystem, while the light quark and gluon cloud probe its color field. Their spectroscopy, electromagnetic structure, and strong, weak, and radiative decays consequently provide complementary tests of confinement dynamics in a three-quark system.

Experimentally, the first evidence for a doubly charmed baryon was reported by the SELEX Collaboration in the $\Xi_{cc}^+\to\Lambda_c^+K^-\pi^+$ channel, with a measured mass of $3518.9\pm0.9$ MeV~\cite{SELEX:2002wqn,ParticleDataGroup:2026mpi}. This signal was not confirmed by subsequent experiments~\cite{Belle:2006edu,BaBar:2006bab,Ratti:2003ez}. A decisive advance was made by the LHCb Collaboration, which observed $\Xi_{cc}^{++}$ in the $\Lambda_c^+K^-\pi^+\pi^+$ invariant-mass spectrum and measured its mass to be $3621.40\pm0.72\pm0.27\pm0.14$ MeV~\cite{LHCb:2017iph}. More recently, LHCb reported a $\Xi_{cc}^{+}$ candidate in the $\Lambda_c^+K^-\pi^+$ final state with a significance above seven standard deviations and a mass of $3619.97\pm0.83\pm0.26^{+1.90}_{-1.30}$ MeV. The resulting isospin splitting, $M_{\Xi_{cc}^{+}}-M_{\Xi_{cc}^{++}}=-1.77\pm0.84\pm0.15^{+1.90}_{-1.30}$ MeV, is consistent with the expected small isospin-breaking effect~\cite{LHCb:2026pxn}. In addition, the LHCb Collaboration reported the observation of $\Omega_{cc}^{+}$ with a measured mass of $M_{\Omega_{cc}^{+}}=3725.9\pm1.0\pm0.2\pm0.4\pm0.6$ MeV~\cite{LHCb:2026lnj}. These developments have inspired theoretical research on doubly charmed baryons.

The mass spectrum of doubly heavy baryons has been studied extensively using constituent-quark and relativistic quark models, bag models, QCD sum rules, lattice QCD, effective field theories and heavy quark symmetry~\cite{Korner:1994nh,Itoh:2000um,Gershtein:2000nx,Ebert:2002ig,Mathur:2002ce,He:2004px,Chiu:2005zc,Brambilla:2005yk,Albertus:2006ya,Zhang:2008rt,Valcarce:2008dr,Hwang:2008dj,Liu:2009jc,Wang:2010vn,Narison:2010py,Brodsky:2011zs,Aliev:2012ru,PACS-CS:2013vie,Wei:2015gsa,Perez-Rubio:2015zqb,Chen:2016spr,Karliner:2017qjm,Chen:2017sbg,Wang:2017qvg,Wang:2017azm,Ghalenovi:2022dok,Yu:2022lel,Pan:2025lmd,Jakhad:2025zos,ShekariTousi:2024mso}. The observed $\Xi_{cc}^{++}$ mass is broadly compatible with the conventional $S$-wave $J^P=1/2^+$ assignment. Its spin-$3/2$ partner $\Xi_{cc}^{*}$ and the strange partners $\Omega_{cc}^{(*)}$ are particularly important because their hyperfine splittings determine whether the strong decay channels are kinematically closed. When this occurs, the radiative transitions $B^*(3/2^+)\to B(1/2^+)\gamma$ become the principal decay modes and offer a clean means of searching for the spin-$3/2$ states.

Electromagnetic transitions probe the spatial and spin structure of hadrons through the magnetic-dipole ($M1$) and electric-quadrupole ($E2$) form factors. The $M1$ transition is expected to dominate for the ground-state $S$-wave baryons, whereas a nonzero $E2$ form factor is sensitive to deformation, relativistic effects. These transitions have been investigated in the bag model, nonrelativistic and relativistic quark models, chiral perturbation theory, and light-cone QCD sum rules~\cite{Branz:2010pq,Dhir:2013nka,Li:2017pxa,Li:2017cfz,Cui:2017udv,Ozdem:2019zis,Sharma:2010vv,Hackman:1977am,Bernotas:2012nz,Xiao:2017udy,Lu:2017meb,Can:2013zpa,Aliev:2023pwd}. The lattice-QCD study of Ref.~\cite{Bahtiyar:2018vub} extracted the $M1$ and $E2$ form factors for the doubly charmed transitions and found a dominant $M1$ amplitude with a very small $E2$ contribution. It also separated the light- and charm-quark electromagnetic currents, demonstrating that the light-quark sector is important while the charm-quark contribution remains non-negligible. A quantitative understanding of the relative light- and heavy-quark photon-emission mechanisms is therefore required for a reliable interpretation of the transition moments.

In this work, we investigate the masses of the low-lying doubly heavy baryons and their $B^*(3/2^+)\to B(1/2^+)\gamma$ radiative transitions within QCD sum rules and LCSR. We calculate the invariant form factors $G_1(0)$ and $G_2(0)$, construct the physical multipole form factors $G_{M1}(0)$ and $G_{E2}(0)$, and determine the transition magnetic moments and radiative widths. Particular attention is paid to the relative light- and heavy-quark contributions, which can significantly affect the magnitude and sign of the $M1$ transition form factor.

This paper is organized as follows. In Sec.~\ref{sec:current}, we use the interpolating currents and perform the QCD sum-rule analysis for the masses of the low-lying doubly heavy baryons. In Sec.~\ref{sec:transitions}, we formulate the electromagnetic transition matrix element, derive the light-cone sum rules for the relevant invariant form factors, and obtain the multipole form factors, transition magnetic moments, and radiative widths. Section~\ref{sec:include} presents the numerical results, comparisons with lattice QCD and other approaches. The explicit operator product expansion (OPE) expressions are collected in Appendix~\ref{app:ope}, while Appendix~\ref{app:qm} gives the corresponding nonrelativistic quark-model calculation of the transition magnetic moments.

\section{Mass analyses from QCD sum rules}
\label{sec:current}

Following Ref.~\cite{Chen:2017sbg}, we systematically examine local interpolating currents for $S$-wave doubly heavy baryons. We consider a ground-state system $Q_1Q_2q$, where $Q_1$ and $Q_2$ are heavy quarks and $q$ is a light quark ($u$, $d$, or $s$). In the absence of orbital angular momentum, the total angular momentum is obtained by coupling the three quark spins:
\begin{equation}
	J=s_{Q_1}\otimes s_{Q_1}\otimes s_{q}= s_q\otimes j_l=s_{Q_1}\otimes j_2=s_{Q_2}\otimes j_3 \, ,
\end{equation}
where $s_{Q_1}$, $s_{Q_2}$, and $s_q$ denote the spins of $Q_1$, $Q_2$, and $q$, respectively. The intermediate angular momenta $j_l$, $j_2$, and $j_3$ correspond to the heavy-heavy, $Q_2q$, and $Q_1q$ subsystems, respectively, and are defined as
\begin{equation}
	j_l=s_{Q_1}\otimes s_{Q_2} \, ,
	j_2=s_{q}\otimes s_{Q_2} \, ,
	j_3=s_{Q_1}\otimes s_{q} \, .
\end{equation}
For two identical heavy quarks, the wave function must be antisymmetric under their exchange.

We consider three types of interpolating currents for doubly heavy baryons, with the general forms

\begin{eqnarray}
	\mathcal{B}_1(x)&=&\epsilon^{abc}[Q_a^T(x)\mathcal{C}\Gamma_1 Q_b(x)]\Gamma_2 q_c(x)\, ,
	\\
	\mathcal{B}_2(x)&=&\epsilon^{abc}[q_a^T(x)\mathcal{C}\Gamma_3 Q_b(x)]\Gamma_4 Q_c(x)\, ,
	\\
	\mathcal{B}_3(x)&=&\epsilon^{abc}[Q_a^T(x)\mathcal{C}\Gamma_5 q_b(x)]\Gamma_6 Q_c(x)\, .
\end{eqnarray}

where $a$, $b$, and $c$ are color indices, with summation over repeated indices implied. The matrix $\mathcal{C}=i\gamma_2\gamma_0$ is the charge-conjugation matrix, the superscript $T$ denotes transposition of the quark fields, and $\Gamma_{1\cdots 6}$ are Dirac matrices specifying the Lorentz structures. The diquark operators with $\Gamma=1/\gamma_5/\gamma_\mu/\gamma_\mu\gamma_5/\sigma_{\mu\nu}$ have quantum numbers $J^P=0^-/0^+/1^+/1^-/1^{\pm }$, respectively. The current $\mathcal{B}_1(x)$ contains a $[QQ]$ diquark whose flavor wave function is symmetric when the heavy quarks are identical. The currents $\mathcal{B}_2(x)$ and $\mathcal{B}_3(x)$ contain heavy-light diquarks and can be transformed into one another.

The doubly heavy baryon currents with $J^P=1/2^+$ and $3/2^+$ were systematically constructed in Ref.~\cite{Chen:2017sbg}. For the $\mathcal{B}_1(x)$ structure, the nonvanishing $J^P=1/2^+$ currents are

\begin{eqnarray}
	\eta_1(x) &=& \epsilon^{abc}[Q_a^T(x)\mathcal{C}\gamma_\mu Q_b(x)]\gamma^\mu\gamma_5 q_c(x)\, ,
	\\
	\eta_2(x) &=& \epsilon^{abc}[Q_a^T(x)\mathcal{C}\sigma_{\mu\nu} Q_b(x)]\sigma^{\mu\nu}\gamma_5 q_c(x)\, ,
\end{eqnarray}
and the nonvanishing $J^P=3/2^+$ currents are
\begin{align}
	\eta_{3\alpha}(x)&=\Gamma_{\alpha \mu}\epsilon^{abc}[Q_a^T(x)\mathcal{C}\gamma_\mu Q_b(x)]q_c(x)\, ,
	\\ \nonumber
	\eta_{4\alpha}(x)&=\Gamma_{\alpha\mu}\epsilon^{abc}\Big{\{}[Q_a^T(x)\mathcal{C}\sigma_{\mu\nu} Q_b(x)]\gamma^{\nu} q_c(x)
	\\ 
	&~~~~+[Q_a^T(x)\mathcal{C}\sigma^{\mu\nu}\gamma_5 Q_b(x)]\gamma_\nu\gamma_5 q_c(x)\Big{\}}\, ,
	\\ \nonumber
	\eta_{5\alpha_1\alpha_2}(x)&=\Gamma_{\alpha_1\alpha_2\mu\nu}\epsilon^{abc}\Big\{[Q_a^T(x)\mathcal{C}\sigma^{\mu\nu} Q_b(x)]\gamma_5 q_c(x)
	\\ 
	&~~~~+[Q_a^T(x)\mathcal{C}\sigma^{\mu\nu}\gamma_5 Q_b(x)] q_c(x)\Big\}\, .
\end{align}

For the $\mathcal{B}_2(x)$ structure, the nonvanishing $J^P=1/2^+$ currents are
\begin{eqnarray}
	\eta_1^\prime(x)&=& \epsilon^{abc}[q_a^T(x)\mathcal{C}\gamma_5 Q_b(x)]Q_c(x)\, ,
	\\
	\eta_2^\prime(x)&=& \epsilon^{abc}[q_a^T(x)\mathcal{C}\gamma_\mu Q_b(x)]\gamma^\mu \gamma_5 Q_c(x)\, ,
	\\
	\eta_6^\prime(x)&=& \epsilon^{abc}[q_a^T(x)\mathcal{C} Q_b(x)]\gamma_5 Q_c(x)\, ,
	\\
	\eta_7^\prime(x)&=& \epsilon^{abc}[q_a^T(x)\mathcal{C}\gamma_\mu\gamma_5 Q_b(x)]\gamma^\mu Q_c(x)\, ,
	\\
	\eta_8^\prime(x)&=& \epsilon^{abc}[q_a^T(x)\mathcal{C}\sigma_{\mu\nu} Q_b(x)]\sigma^{\mu\nu}\gamma_5 Q_c(x)\, ,
\end{eqnarray}

The corresponding $J^P=3/2^+$ currents comprise two Rarita--Schwinger fields and one tensor field:
\begin{align}
	\eta_{3\alpha}^\prime(x)&=\Gamma_{\alpha \mu}\epsilon^{abc}[q_a^T(x)\mathcal{C}\gamma_\mu Q_b(x)]Q_c(x)\, ,
	\\ \nonumber
	\eta_{4\alpha}^\prime(x)&=\Gamma_{\alpha\mu}\epsilon^{abc}\Big{\{}[q_a^T(x)\mathcal{C}\sigma_{\mu\nu} Q_b(x)]\gamma^{\nu} Q_c(x)
	\\
	&~~~~+[q_a^T(x)\mathcal{C}\sigma^{\mu\nu}\gamma_5 Q_b(x)]\gamma_\nu\gamma_5 Q_c(x)\Big{\}}\, ,
	\\ \nonumber
	\eta_{5\alpha_1\alpha_2}^\prime(x)&=\Gamma_{\alpha_1\alpha_2\mu\nu}\epsilon^{abc}\Big\{[q_a^T(x)\mathcal{C}\sigma^{\mu\nu} Q_b(x)]\gamma_5 Q_c(x)
	\\
	&~~~~+[q_a^T(x)\mathcal{C}\sigma^{\mu\nu}\gamma_5 Q_b(x)] Q_c(x)\Big\}\, ,
\end{align}

The projection operators appearing in these currents are defined by
\begin{align}
	\Gamma_{\mu\nu}&=g_{\mu\nu}-{1\over4}\gamma_\mu\gamma_\nu\, ,
	\\ \nonumber \Gamma_{\mu\nu\alpha_1\alpha_2}&=g_{\mu\alpha_1}g_{\nu\alpha_2}+{1\over 2}g_{\mu\alpha_2}\gamma_\nu\gamma_{\alpha_2}-{1\over2}g_{\nu\alpha_2}\gamma_\mu\gamma_{\alpha_1}
	\\ &~~~~+{1\over 6}\sigma_{\mu\nu}\sigma_{\alpha_1\alpha_2}\, .
\end{align}

Fierz transformations relate $\eta_1^\prime$, $\eta_2^\prime$, $\eta_6^\prime$, $\eta_7^\prime$, and $\eta_8^\prime$ to $\eta_1$ and $\eta_2$, leaving only two independent currents in the $J^P=1/2^+$ sector. Similarly, $\eta_{3\alpha}^\prime$ and $\eta_{4\alpha}^\prime$ can be related to $\eta_{3\alpha}$ and $\eta_{4\alpha}$ through Fierz transformations.

We use these currents to study the ground-state mass spectra of $\Xi_{cc/bb}$ and $\Omega_{cc/bb}$. Specifically, we employ $\eta_1$ and $\eta_1^\prime$ for the $J^P=1/2^+$ states, and $\eta_{3\alpha}$ and $\eta_{3\alpha}^\prime$ for their $J^P=3/2^+$ partners. Taking $\Omega_{cc}(1/2^+)$ as an example, we define its coupling to the interpolating current through
\begin{eqnarray}
	\langle 0|\eta|\Omega_{cc}(1/2^+)\rangle=f_{\Omega_{cc}} u(p)\, .
\end{eqnarray}
where $f_{\Omega_{cc}}$ is the current--baryon coupling, referred to as the decay constant, and $u(p)$ is the Dirac spinor. To construct the QCD sum rules, we evaluate the two-point correlation function at both the hadronic and quark--gluon levels:
\begin{eqnarray}
	\Pi(q^2)=i\int d^4 x e^{iq\cdot x}\langle 0|\eta(x)\bar{\eta}(0)|0\rangle=(q\!\!\!\slash +M_{\Omega_{cc}})\Pi(q^2)\, .
\end{eqnarray}

At the hadronic level, the invariant amplitude $\Pi(q^2)$ is expressed through the dispersion relation
\begin{eqnarray}
	\Pi(q^2)=\int_{m_s^2}^\infty {\rho(s)\over s-q^2-i\epsilon} ds\, .
\end{eqnarray}

where $\rho(s)$ is the spectral density. It a ground-state pole plus contributions from higher resonances and the continuum:
\begin{eqnarray}
	\nonumber \rho(s)&=&\sum_n\delta(s-M_n)\langle 0|\eta|B_n\rangle\langle B_n|\bar\eta|0\rangle
	\\
	&=& f_{\Omega_{cc}}^2\delta(s-M_{\Omega_{cc}}^2)(q\!\!\!\slash +M_{\Omega_{cc}})+\cdots\, .
\end{eqnarray}

At the quark--gluon level, we insert the interpolating currents $\eta_1$ and $\eta_1^\prime$ into the two-point correlation function and evaluate it using the operator product expansion. After applying a Borel transformation and subtracting the continuum contribution above the threshold $s_0$, we match the hadronic and OPE representations to obtain
\begin{align}
	\label{eq:Borel}\Pi(M_B^2,s_0)=f_{\Omega_{cc}}^2 M_{\Omega_{cc}} e^{-M_{\Omega_{cc}}^2/M_B^2}=\int_{m_s^2}^{s_0} e^{-s/M_B^2} \rho(s) ds\, .
\end{align}

where $M_B$ is the Borel mass and $s_0$ is the continuum threshold. Differentiating Eq.~(\ref{eq:Borel}) with respect to $-1/M_B^2$ and dividing by the original sum rule, we obtain the mass of $\Omega_{cc}(1/2^+)$:
\begin{eqnarray}
	M_{\Omega_{cc}}^2={{\partial\over \partial(-1/M_B^2)}\Pi(M_B^2,s_0)\over \Pi(M_B^2,s_0)}={\int_{m_s^2}^{s_0} e^{-s/M_B^2}s \rho(s) ds\over \int_{m_s^2}^{s_0} e^{-s/M_B^2} \rho(s) ds }\, .
\label{eq:mass}
\end{eqnarray}

The OPE calculation of Eq.~(\ref{eq:Borel}) yields the spectral densities associated with $\Pi_{\eta_1^\prime\eta_1^\prime}$, $\Pi_{\eta_1\eta_1}$, and $\Pi_{\eta_1^\prime\eta_1}$. Their explicit expressions are given in Appendix~\ref{app:ope}.

For the numerical analysis, we adopt the following quark masses and condensate parameters:
\begin{eqnarray}
	\nonumber &&\langle \bar q q\rangle=-(0.24\pm 0.01)^3\rm{GeV^3}\, ,
	\\
	\nonumber &&\langle \bar s s\rangle=0.8\times \langle \bar q q\rangle\, ,
	\\
	\nonumber &&\langle g_s^2 GG\rangle=(0.48\pm 0.14)\rm{GeV^4}\, ,
	\\
	&&\langle g_s \bar q\sigma Gq \rangle=M_0^2\times \langle \bar q q\rangle\, ,
	\\
	\nonumber &&\langle g_s \bar s\sigma Gs \rangle=M_0^2\times \langle \bar s s\rangle\, ,
	\\
	\nonumber && M_0^2=0.8\rm{GeV^2}\, ,
	\\
	\nonumber && m_s(2\rm{GeV})=96^{+8}_{-4}\rm{MeV}\, ,
	\\
	\nonumber && m_c=1.23\pm 0.09 \rm{GeV}\, ,
	\\
	\nonumber && m_d=4.18^{+0.04}_{-0.03}\rm{GeV}\, .
\end{eqnarray}

The mass extracted from Eq.~(\ref{eq:mass}) depends on the continuum threshold $s_0$ and the Borel mass $M_B$. We constrain their working ranges using three criteria: convergence of the OPE, dominance of the pole contribution, and stability of the extracted mass. For OPE convergence, we require the contributions of dimension-$8$ and dimension-$10$ to be smaller than $1\%$ each. We also require a pole contribution (PC) larger than $80\%$ and a sufficiently weak dependence of the mass on both parameters. The convergence ratios and pole contribution are defined as follows:
\begin{eqnarray}
	\rm{CVG}\equiv|{\Pi^{\rm{Dim=8}}(M_B^2,\infty)\over \Pi(M_B^2,\infty)}|< 1\%\, ,
	\\
	\rm{CVG}^\prime \equiv|{\Pi^{\rm{Dim=10}}(M_B^2,\infty)\over \Pi(M_B^2,\infty)}|< 1\%\, ,
	\\
	\rm{PC}\equiv|{\Pi(M_B^2,s_0)\over \Pi(M_B^2,\infty)}|> 80\%\, ,
\end{eqnarray}

We first analyze the diagonal correlators $\Pi_{\eta_1^\prime\eta_1^\prime}$ and $\Pi_{\eta_1\eta_1}$, together with the off-diagonal correlator $\Pi_{\eta_1^\prime\eta_1}$. For $\Pi_{\eta_1^\prime\eta_1^\prime}$, a reliable QCD sum-rule result is obtained in the ranges $22\rm{GeV^2}<s_0<28\rm{GeV^2}$ and $2.82\rm{GeV^2}<M_B^2<3.42\rm{GeV^2}$. At the central values $s_0=25\rm{GeV^2}$ and $M_B^2=3.12\rm{GeV^2}$, the extracted mass of $\Omega_{cc}$ is $3.88 \rm{GeV}$. For $\Pi_{\eta_1\eta_1}$, the perturbative and mixed quark--gluon condensate contributions are absent, so $\eta_1$ alone does not provide a reliable sum rule. Moreover, the ratio of $|\Pi_{\eta_1^\prime\eta_1}|$ to $|\Pi_{\eta_1^\prime\eta_1^\prime}|$ is approximately $2$ within the Borel window, indicating a substantial off-diagonal contribution. We therefore introduce the mixed currents $J_1$ and $J_2$ with $\theta_2=-55^\circ$:
\begin{eqnarray}
	J_1=\cos\theta \eta_1^\prime +\sin\theta \eta_1\, ,
	\\
	J_2=-\sin\theta \eta_1^\prime +\cos\theta \eta_1\, .
\end{eqnarray}
We then perform QCD sum rule analyses with $J_1$ and $J_2$. The correlator $\Pi_{J_1J_1}$ yields a physically acceptable result, whereas $\Pi_{J_2 J_2}$ does not. For $\Pi_{J_1J_1}$, we choose the working ranges $22 \rm{GeV^2}<s_0<28 \rm{GeV^2}$ and $2.82 \rm{GeV^2}<M_B^2<3.42 \rm{GeV^2}$ and obtain
\begin{eqnarray}
M_{\Omega_{cc}(1/2^+)}&=&3.72^{+0.13}_{-0.15}~\rm{GeV}\, ,
\\ 
f_{\Omega_{cc}(1/2^+)}&=&0.196^{+0.023}_{-0.030}~\rm{GeV^3}\, .
\end{eqnarray}

\begin{figure*}[hbtp]
\begin{center}
\subfigure[]{
\scalebox{0.6}{\includegraphics{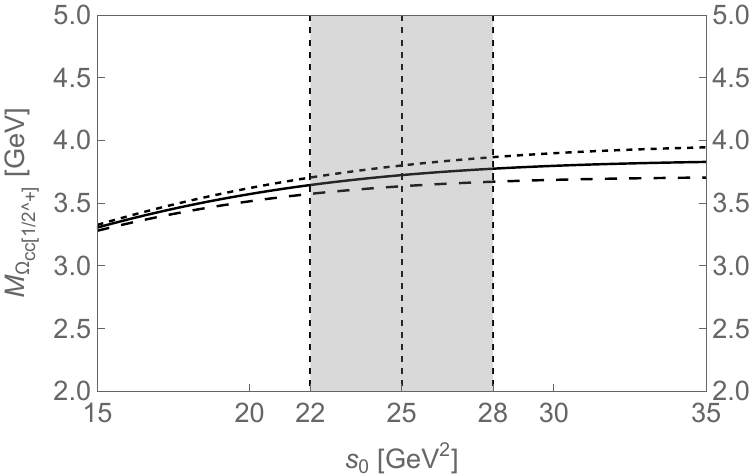}}}~~~
\subfigure[]{
\scalebox{0.6}{\includegraphics{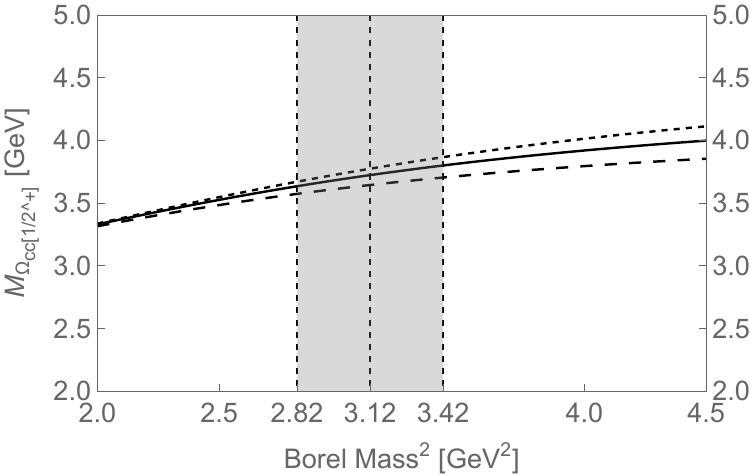}}}
\end{center}
\caption{Dependence of the extracted $\Omega_{cc}$ mass on the continuum threshold $s_0$ (left) and the Borel parameter $M_B^2$ (right). In the left panel, the long-dashed, solid, and short-dashed curves correspond to $M_B^2=2.82~\rm{GeV^2}$, $3.12~\rm{GeV^2}$, and $3.42~\rm{GeV^2}$, respectively. In the right panel, the same line styles correspond to $s_0=22~\rm{GeV^2}$, $25~\rm{GeV^2}$, and $28~\rm{GeV^2}$, respectively.}
\label{fig:mass}
\end{figure*}

\begin{figure*}[hbtp]
\begin{center}
\subfigure[]{
\scalebox{0.5}{\includegraphics{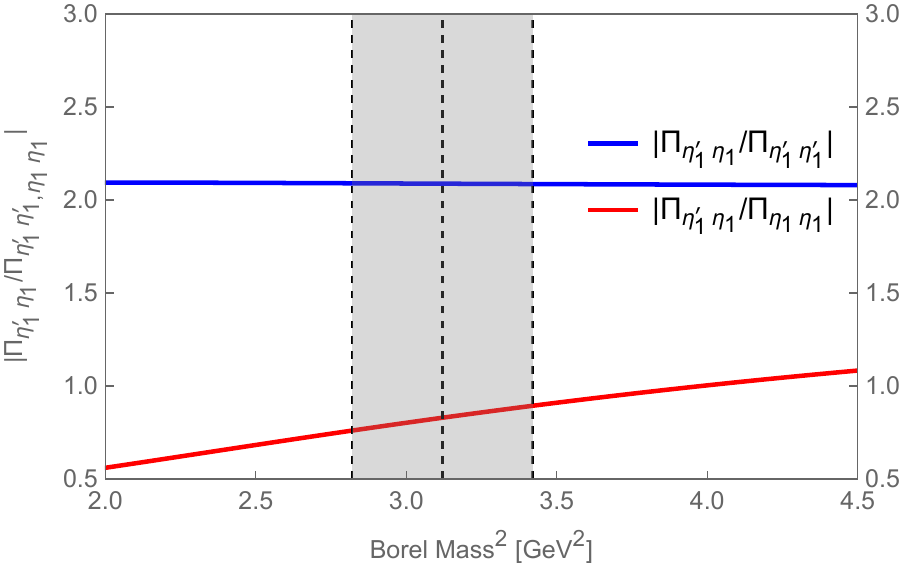}}}~~~
\subfigure[]{
\scalebox{0.6}{\includegraphics{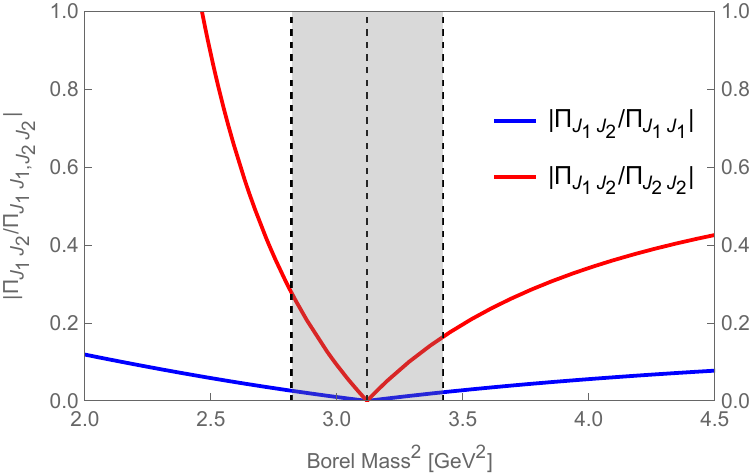}}}
\end{center}
\caption{Relative off-diagonal contributions as functions of the Borel parameter $M_B^2$. The left panel shows $|\Pi_{\eta_1^\prime\eta_1}/\Pi_{\eta_1^\prime\eta_1^\prime}|$ and $|\Pi_{\eta_1^\prime\eta_1}/\Pi_{\eta_1\eta_1}|$ before current mixing. The right panel shows $|\Pi_{J_1 J_2}/\Pi_{J_1 J_1}|$ and $|\Pi_{J_1 J_2}/\Pi_{J_2 J_2}|$ after current mixing.}
\label{fig:ratio}
\end{figure*}

Fig.~\ref{fig:mass} shows the dependence of the extracted $\Omega_{cc}$ mass on $s_0$ and $M_B^2$ in the left and right panels, respectively. Fig.~\ref{fig:ratio} compares the relative off-diagonal contributions before and after current mixing. The left panel shows $|\Pi_{\eta_1^\prime\eta_1}/\Pi_{\eta_1^\prime\eta_1^\prime}|$ and $|\Pi_{\eta_1^\prime\eta_1}/\Pi_{\eta_1\eta_1}|$ as functions of $M_B^2$. While the right panel shows the ratios of $|\Pi_{J_1 J_2}/\Pi_{J_1 J_1}|$ and $|\Pi_{J_1 J_2}/\Pi_{J_2 J_2}|$. The right panel indicates a small residual off-diagonal contribution between $J_1$ and $J_2$ relative to $J_1$ in the range $2.82 \rm{GeV^2}<M_B^2<3.42 \rm{GeV^2}$.

We apply the same procedure to $J^P=3/2^+$ system, mixing $\eta_{3\mu}^\prime$ and $\eta_{3\mu}$ with an angle $\theta^\prime_i$ to construct $J_{1\mu}$ and $J_{2\mu}$. Table~\ref{tab:results} summarizes the parameters and results for all baryons considered.

\begin{table*}[hbtp]
	\begin{center}
		\renewcommand{\arraystretch}{1.6}
		\caption{QCD sum-rule parameters and results for the ground-state doubly heavy baryons. The columns list the interpolating currents, mixing angles, baryons, continuum thresholds, Borel windows, masses, mass differences between spin partners, and decay constants.}
		\begin{tabular}{ c  c  c  c  c  c  c  c}
			\hline\hline
			Currents                   &~Mixing Angle                    & ~Baryons       & ~$s_0[\rm{GeV^2}]$ &~ $M_B^2[\rm{GeV^2}]$~ & Mass[$\rm{GeV}$]~        & Difference[$\rm{MeV}$]  ~            & Decay constant[$\rm{GeV^3}$]
			\\ \hline
			$J_1^{\Xi_{cc}}$           & $\theta_1=-54\pm 5^\circ$       & $\Xi_{cc}$      & $25\pm3$          & $2.70<M_B^2<3.30$   & $3.62^{+0.14}_{-0.15}$  & \multirow{2}{*}{$101^{+27}_{-11}$}    & $0.171^{+0.021}_{-0.025}$
			\\
			$J_{1\mu}^{\Xi_{cc}^*}$    & $\theta_1^\prime=35\pm 5^\circ$ & $\Xi_{cc}^*$    & $25\pm3$          & $2.54<M_B^2<3.14$   & $3.72^{+0.14}_{-0.13}$  &                                       & $0.097 ^{+0.013}_{-0.015}$
			\\
			$J_1^{\Omega_{cc}}$        & $\theta_2=-55\pm 5^\circ$       & $\Omega_{cc}$   & $25\pm3$          & $2.82<M_B^2<3.42$   & $3.72^{+0.13}_{-0.15}$  & \multirow{2}{*}{$99^{+11}_{-12}$}     & $0.196^{+0.023}_{-0.030}$
			\\
			$J_{1\mu}^{\Omega_{cc}^*}$ & $\theta_2^\prime=37\pm 5^\circ$ & $\Omega_{cc}^*$ & $25\pm3$          & $2.84<M_B^2<3.44$   & $3.82^{+0.13}_{-0.15}$  &                                       & $0.112^{+0.014}_{-0.018}$
			\\
			$J_1^{\Xi_{bb}}$           & $\theta_3=-44\pm5^\circ$        & $\Xi_{bb}$      & $140\pm10$        & $11.15<M_B^2<13.15$ & $10.27^{+0.15}_{-0.19}$ & \multirow{2}{*}{$30^{+18}_{-17}$} & $0.959^{+0.144}_{-0.175}$
			\\
			$J_1^{\Xi_{bb}^*}$         & $\theta_3^\prime=27\pm5^\circ$  & $\Xi_{bb}^*$    & $140\pm10$        & $10.57<M_B^2<12.57$ & $10.30^{+0.14}_{-0.18}$ &                                       & $0.566^{+0.082}_{-0.102}$
			\\
			$J_1^{\Omega_{bb}}$        & $\theta_4=-47\pm5^\circ$        & $\Omega_{bb}$   & $140\pm10$        & $11.53<M_B^2<13.53$ & $10.32^{+0.15}_{-0.19}$ & \multirow{2}{*}{$21^{+14}_{-13}$}     & $1.060^{+0.157}_{-0.190}$
			\\
			$J_{1\mu}^{\Omega_{bb}^*}$ & $\theta_4^\prime=29\pm5^\circ$  & $\Omega_{bb}^*$ & $140\pm10$        & $10.97<M_B^2<12.97$ & $10.34^{+0.15}_{-0.18}$ &                                       & $0.612^{+0.089}_{-0.110}$
			\\ \hline\hline
		\end{tabular}
		\label{tab:results}
	\end{center}
\end{table*}

\section{Electromagnetic transitions from light-cone sum rules}
\label{sec:transitions}

In this section, we study the radiative transitions of ground-state doubly heavy baryons, $B^*(3/2^+)\to B(1/2^+)+\gamma$, using light-cone sum rules. We first introduce the electromagnetic transition matrix element and its covariant decomposition, and then express the radiative decay width and transition magnetic moment in terms of the magnetic-dipole and electric-quadrupole form factors, $G_{M1}$ and $G_{E2}$. We subsequently derive the light-cone sum rules and present the numerical results.

For the process
\begin{equation}
	B^*(p)\to B(k)+\gamma(q,\epsilon),\qquad p=k+q,
\end{equation}
we express the electromagnetic transition matrix element as
\begin{equation}
	\langle B^*(p)|B(k)\gamma(q,\epsilon)\rangle
	= e\,\bar u^{\rho}(p)\,O_{\rho\mu}\,u(k)\,\epsilon^{\mu},
\end{equation}
where $B^*$ and $B$ denote the ground-state baryons of $J=3/2$ and $J=1/2$, respectively. The Rarita--Schwinger spinor $u^{\rho}(p)$ satisfies the on-shell constraints
\begin{equation}
	p^{\rho}u_{\rho}(p)=0,\qquad \gamma^{\rho}u_{\rho}(p)=0,
\end{equation}
The photon polarization vector is denoted by $\epsilon^{\mu}$.

We decompose the transition tensor $O_{\rho\mu}$ into two independent Lorentz structures:
\begin{eqnarray}
\label{eq:Orhomu}
	O_{\rho\mu}&=&
	\frac{G_1}{2M_B}(q_{\rho}\gamma_{\mu}-\slashed q\,g_{\rho\mu})\gamma_5
	\\ 
\nonumber &&+\frac{G_2}{4M_B^2}\frac{1}{M_B+M_B^*}(q\!\cdot\!p\,g_{\rho\mu}-q_{\rho}p_{\mu})\slashed q\,\gamma_5,
\end{eqnarray}
where $M_B$ and $M_B^*$ are the masses of the final $1/2^+$ and initial $3/2^+$ baryons, respectively. Each Lorentz structure satisfies the electromagnetic gauge-invariance condition $q^{\mu}O_{\rho\mu}=0$, as follows
\begin{eqnarray}
	q^{\mu}(q_{\rho}\gamma_{\mu}-\slashed q\,g_{\rho\mu})=q_{\rho}\slashed q-\slashed q\,q_{\rho}=0,
\end{eqnarray}
and
\begin{eqnarray}
	q^{\mu}(q\!\cdot\!p\,g_{\rho\mu}-q_{\rho}p_{\mu})=q_{\rho}(q\!\cdot\!p)-q_{\rho}(q\!\cdot\!p)=0.
\end{eqnarray}

The factor $\slashed q$ in the $G_2$ term distinguishes its Dirac structure from that of the $G_1$ term, allowing the two form factors to be projected separately in the LCSR calculation. For a real photon and on-shell external baryons, the transition can be expressed in terms of magnetic-dipole and electric-quadrupole multipoles, as discussed in Sec.~\ref{sec:multipoles}.

\subsection{Multipole form factors and electromagnetic transitions}
\label{sec:multipoles}

The magnetic-dipole and electric-quadrupole form factors are related to $G_1$ and $G_2$ by
\begin{eqnarray}
	G_{M1}&=&\frac{2}{3}G_1-\frac{\delta}{6M_B^*}G_1-\frac{\delta}{12M_B}G_2\, ,
	\\ G_{E2}&=&\frac{\delta}{6M_B^*}G_1-\frac{\delta}{12M_B}G_2,
\end{eqnarray}
where $\delta=|\mathbf q|=(M_B^{*2}-M_B^2)/(2M_B^*)$ is the three-momentum of the photon in the rest frame of $B^*$.

The invariant form factors $G_1$ and $G_2$ are not directly observable. The radiative decay width is expressed in terms of the multipole form factors $G_{M1}(0)$ and $G_{E2}(0)$. In most numerical analyses, the $M1$ contribution dominates, while the $E2$ contribution is small.

The two-body decay width is given by
\begin{equation}
	\Gamma=\frac{|\mathbf q|}{8\pi M_B^{*2}}\frac{1}{2J_{B^*}+1}\sum_{\rm spins}|\mathcal M|^2,
\end{equation}
where $J_{B^*}=3/2$ and $|\mathbf q|=(M_B^{*2}-M_B^2)/(2M_B^*)$. In terms of the helicity amplitudes $A_{1/2}$ and $A_{3/2}$, the squared matrix element, averaged over the initial spin and summed over the final polarizations, becomes
\begin{eqnarray}
	\frac{1}{4}\sum_{\rm spins}|\mathcal M|^2 &=&4 |\mathbf q| m_B m_B^*(|A_{1/2}|^2+|A_{3/2}|^2)\\ \nonumber
	&=&\frac{4\pi\alpha\,M_B^{*2}|\mathbf q|^2}{M_B^2}\left(|G_{M1}(0)|^2+3|G_{E2}(0)|^2\right).
\end{eqnarray}
The radiative decay width can therefore be written in terms of the multipole form factors as
\begin{equation}
	\Gamma(B^*\to B\gamma)=
	\frac{\alpha}{16}
	\frac{(M_B^{*2}-M_B^2)^3}{M_B^{*3}M_B^2}
	\left(|G_{M1}(0)|^2+3|G_{E2}(0)|^2\right),
	\label{eq:width}
\end{equation}
where $\alpha=e^2/(4\pi)\simeq 1/137$ is the fine-structure constant.

The corresponding transition magnetic moment is given by
\begin{equation}
	\mu(B^*\to B\gamma)=
	\frac{2M_B^*}{M_B^*+M_B}\,G_{M1}(0)\,\frac{e}{2M_B}.
	\label{eq:mu}
\end{equation}
Thus, determining $G_1(0)$ and $G_2(0)$ from the light-cone sum rules allows us to calculate $G_{M1}(0)$ and $G_{E2}(0)$, and hence the transition magnetic moment and radiative decay width.

\subsection{Light-cone sum rules}
\label{sec:lcsr}

To derive the light-cone sum rules for $B^*\to B\gamma$, we consider the two-point correlation function in an external photon state:
\begin{equation}
	\Pi_{\alpha}(p,k,q,\epsilon)=
	\int d^4x\, e^{-ik\cdot x}
	\langle 0|J_{B^*,\alpha}(0)\bar J_B(x)|\gamma(q,\epsilon)\rangle,
	\label{eq:correlator}
\end{equation}
where $J_{B^*,\alpha}$ and $J_B$ are the interpolating currents constructed in Sec.~\ref{sec:current} for the $3/2^+$ and $1/2^+$ doubly heavy baryons, respectively.

At the hadronic level, we insert complete sets of intermediate states and define the pole residues through
\begin{eqnarray}
	\langle 0|J_B(0)|B\rangle&=&f_B\,u_B(k),
	\\ \nonumber \langle 0|J_{B^*,\alpha}(0)|B^*\rangle&=&f_{B^*}\,u_{B^*,\alpha}(p).
\end{eqnarray}
The ground-state contribution to the correlation function is then
\begin{equation}
	\Pi_{\alpha}(p,k,q,\epsilon)=
	\frac{f_{B^*}f_B\,u_{\alpha}(p)\,\langle B^*(p)|B(k)\gamma(q,\epsilon)\rangle\,\bar u(k)}
	{(p^2-M_B^{*2})(k^2-M_B^2)}.
\end{equation}
After substituting Eq.~(\ref{eq:Orhomu}) and we decompose the correlation function into independent Lorentz structures.

To extract $G_1$ and $G_2$ separately, we select the following Lorentz structures:
\begin{eqnarray}
	\mathcal T_1&=&\epsilon_{\alpha\nu\rho\sigma}\,\epsilon^{\nu}q^{\rho}p^{\sigma},
	\\ \mathcal T_2&=&\sigma^{\mu\nu}\epsilon_{\mu\nu\rho\sigma}p^{\rho}q^{\sigma}q^{\alpha}\,\epsilon\!\cdot\!p.
\end{eqnarray}
Denoting the corresponding projected contributions by $\Pi_{\alpha,1}$ and $\Pi_{\alpha,2}$, we obtain the hadronic representations
\begin{align}
\begin{split}
\Pi_{\alpha,1}
&= \frac{e\,f_{B^*}f_B(M_B^*+M_B)}
         {3M_B(p^2-M_B^{*2})(k^2-M_B^2)}G_1
\\
&\quad \times
\epsilon_{\alpha\nu\rho\sigma}
\epsilon^{\nu}q^{\rho}p^{\sigma},
\end{split}
\\
\begin{split}
\Pi_{\alpha,2}
&= \frac{e\,f_{B^*}f_B }
         {8M_B^2(p^2-M_B^{*2})(k^2-M_B^2)}G_2
\\
&\quad \times
\sigma^{\mu\nu}\epsilon_{\mu\nu\rho\sigma}
p^{\rho}q^{\sigma}q^{\alpha}\,
\epsilon\!\cdot\!p.
\end{split}
\end{align}

These projections isolate $G_1$ and $G_2$ individually, without requiring the solution of coupled linear equations.

At the quark--gluon level, we evaluate the correlation function using the light-cone photon distribution amplitudes~\cite{Ball:2002ps}. We illustrate the calculation with the decay $\Omega_{cc}^*(3/2^+)\to \Omega_{cc}(1/2^+)+\gamma$. The explicit spectral densities $\rho_{1}^{(\rm OPE)}(s)$ and $\rho_{2}^{(\rm OPE)}(s)$, including perturbative terms and quark and gluon condensate contributions, are collected in Appendix~\ref{app:ope}.

After applying a Borel transformation and subtracting the continuum contribution, we match the hadronic and OPE representations to obtain the sum rules for $G_1$ and $G_2$:
\begin{equation}
	\label{eq:sumG1} \frac{2e\,f_{B^*}f_B}{3}
	G_1\frac{e^{-M^2/M_B^2}}{M_B^2}=
 \int_{s_<}^{s_0} e^{-s/M_B^2}\bigl(\rho_{1}^{l}(s)+\rho_{1}^{h}(s)\bigr)ds,
\end{equation}
\begin{equation}
	\label{eq:sumG2}
\frac{e\,f_{B^*}f_B}{8M^2}
	G_2\frac{e^{-M^2/M_B^2}}{M_B^2}=
\int_{s_<}^{s_0} e^{-s/M_B^2}\bigl(\rho_{2}^{l}(s)+\rho_{2}^{l}(s)\bigr)ds,
\end{equation}
Here, $s_0$ is the continuum threshold and $M_B$ denotes the Borel mass. In deriving these sum rules, we use the approximations and average baryon mass defined by
\begin{equation}
	p\approx k\gg q,\qquad M_B^*\approx M_B,\qquad M=\frac{M_B^*+M_B}{2}.
\end{equation}

The spectral densities contain two types of contributions. The light-quark contribution, $\rho^{l}(s)$, describes photon couplings through the light-cone distribution amplitudes, whereas the heavy-quark contribution, $\rho^{h}(s)$, describes direct photon emission from heavy-quark lines. We extract $G_1(0)$ and $G_2(0)$ numerically from Eqs.~(\ref{eq:sumG1}) and (\ref{eq:sumG2}) and use the multipole relations to determine $G_{M1}(0)$ and $G_{E2}(0)$. The transition magnetic moment $\mu$ and radiative width $\Gamma$ then follow from Eqs.~(\ref{eq:width})--(\ref{eq:mu}). 

\begin{figure*}[hbtp]
\begin{center}
\subfigure[]{
\scalebox{0.63}{\includegraphics{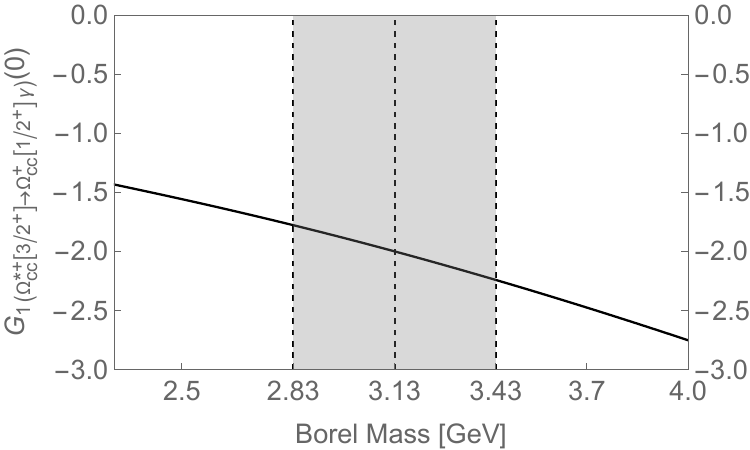}}}~~~
\subfigure[]{
\scalebox{0.6}{\includegraphics{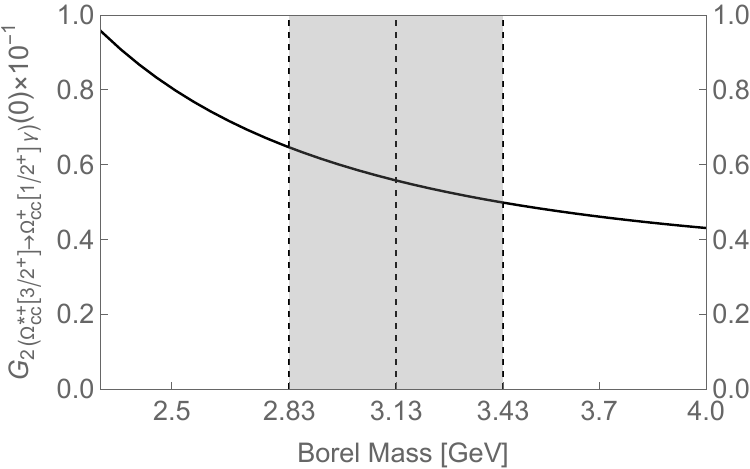}}}
\end{center}
\caption{Dependence of the form factors $G_1(0)$ (left) and $G_2(0)$ (right) on the Borel parameter $M_B^2$.}
\label{fig:form factor}
\end{figure*}

For the $\Omega_{cc}^*(3/2^+)$ and $\Omega_{cc}(1/2^+)$ states, we use the baryon masses listed in Table~\ref{tab:results} and set $s^0=25~\rm{GeV}^2$ and $M_B^2=3.13~\rm{GeV}^2$. These choices correspond to the averages of the continuum thresholds and central Borel parameters used in the respective mass sum rules. Fig.~\ref{fig:form factor} shows $G_1(0)$ and $G_2(0)$ as functions of the Borel parameter $M_B^2$. The light-cone sum rules yield the form factors $G_1(0)$ and $G_2(0)$:
\begin{eqnarray}
	G_{1,\Omega_{cc}^*\to \Omega_{cc}\gamma}(0)&=&-1.999^{+0.890}_{-1.232}\, ,
	\\ G_{2,\Omega_{cc}^*\to \Omega_{cc}\gamma}(0)&=&0.056^{+0.034}_{-0.027}\, ,
\end{eqnarray}
The sources of uncertainty include the continuum threshold, the Borel parameter, the $\Omega_{cc}^*$ and $\Omega_{cc}$ parameters, the quark and gluon inputs, and the parameters of the distribution amplitudes. From these form factors, we obtain $G_{M1}(0)$, $G_{E2}(0)$, the transition magnetic moment $\mu$, and the radiative width $\Gamma$ for $\Omega_{cc}^*\to \Omega_{cc}\gamma$:
\begin{eqnarray}
	G_{M1,\Omega_{cc}^*\to \Omega_{cc}\gamma}(0)&=&-1.324^{+0.590}_{-0.816}\, ,
	\\ G_{E2,\Omega_{cc}^*\to \Omega_{cc}\gamma}(0)&=&-0.008^{+0.004}_{+0.005}\, ,
	\\ \mu_{\Omega_{cc}^*\to \Omega_{cc}\gamma}&=&-0.34^{+0.15}_{-0.21}\mu_N
	\\ \Gamma_{\Omega_{cc}^*\to \Omega_{cc}\gamma}&=&0.41^{+0.67}_{-0.30}~\rm{keV}\, .
\end{eqnarray}
The input parameters and numerical results are summarized in Tables~\ref{tab:em_input} and \ref{tab:em_result}, respectively. We use $\delta=(M_B^{*2}-M_B^2)/(2M_B^*)$, $\alpha=1/137$, and $m_N=0.938$~GeV.

Applying the same procedure to all six channels, we obtain the following form factors:
\begin{eqnarray}
	G_{1,\Xi_{cc}^{*++}\to \Xi_{cc}^{++}\gamma}(0)&=&-6.388^{+2.367}_{-3.457}\, ,
	\\ G_{2,\Xi_{cc}^{*++}\to \Xi_{cc}^{++}\gamma}(0)&=&-0.219^{+0.102}_{-0.223}\, ,
	\\
	G_{1,\Xi_{cc}^{*+}\to \Xi_{cc}^{+}\gamma}(0)&=&-1.507^{+0.728}_{-1.369}\, ,
	\\ G_{2,\Xi_{cc}^{*+}\to \Xi_{cc}^{+}\gamma}(0)&=&0.109^{+0.111}_{-0.058}\, ,
  \\G_{1,\Omega_{cc}^*\to \Omega_{cc}\gamma}(0)&=&-1.999^{+0.890}_{-1.232}\, ,
	\\ G_{2,\Omega_{cc}^*\to \Omega_{cc}\gamma}(0)&=&0.056^{+0.034}_{-0.027}\, ,
	\\ G_{1,\Xi_{bb}^{*0}\to \Xi_{bb}^{0}\gamma}(0)&=&-4.490^{+1.554}_{-1.923}\, ,
	\\ G_{2,\Xi_{bb}^{*0}\to \Xi_{bb}^{0}\gamma}(0)&=&-0.157^{+0.075}_{-0.082}\, ,
	\\
	G_{1,\Xi_{bb}^{*-}\to \Xi_{bb}^{-}\gamma}(0)&=&4.883^{+1.922}_{-1.538}\, ,
	\\ G_{2,\Xi_{bb}^{*-}\to \Xi_{bb}^{-}\gamma}(0)&=&0.079^{+0.041}_{-0.038}\, ,
	\\ G_{1,\Omega_{bb}^{*-}\to \Omega_{bb}^{-}\gamma}(0)&=&3.880^{+1.484}_{-1.183}\, ,
	\\ G_{2,\Omega_{bb}^{*-}\to \Omega_{bb}^{-}\gamma}(0)&=&0.041^{+0.022}_{-0.020}\, .
\end{eqnarray}

Combining these form factors with mass parameters listed in Table~\ref{tab:em_input}, we calculate the multipole form factors $G_{M1}(0)$ and $G_{E2}(0)$, transition magnetic moments, and radiative widths for the six ground-state transitions $B^*(3/2^+)\to B(1/2^+)\gamma$
\begin{eqnarray}
	\Gamma_{\Xi_{cc}^{*++}\to \Xi_{cc}^{++}\gamma}&=&5.25^{+8.57}_{-3.62}~\rm{keV}\, ,
	\\ \Gamma_{\Xi_{cc}^{*+}\to \Xi_{cc}^{+}\gamma}&=&0.31^{+0.84}_{-0.25}~\rm{keV}\, ,
	\\ \Gamma_{\Omega_{cc}^{*+}\to \Omega_{cc}^{+}\gamma}&=&0.41^{+0.67}_{-0.30}~\rm{keV}\, ,
	\\ \Gamma_{\Xi_{bb}^{*0}\to \Xi_{bb}^{0}\gamma}&=&0.01^{+0.01}_{-0.01}~\rm{keV}\, ,
	\\ \Gamma_{\Xi_{bb}^{*-}\to \Xi_{bb}^{-}\gamma}&=&0.01^{+0.02}_{-0.01}~\rm{keV}\, ,
	\\ \Gamma_{\Omega_{bb}^{*-}\to \Omega_{bb}^{-}\gamma}&=&0.002^{+0.003}_{-0.002}~\rm{keV}\, .
\end{eqnarray}
The results are summarized in Table~\ref{tab:em_result}.

In the following, we compare the numerical results with lattice QCD and other theoretical approaches.
\begin{table*}[hbtp]
	\begin{center}
		\renewcommand{\arraystretch}{1.6}
	\caption{Baryon masses and invariant form factors used to evaluate the six radiative decay channels.}
	\label{tab:em_input}
	\begin{tabular}{lcccc}
		\hline\hline
		Channel                                    & $M_B^*$ [GeV] & $M_B$ [GeV]      & $G_1(0)$ & $G_2(0)$ \\
		\hline
		$\Xi_{cc}^{*++}\to\Xi_{cc}^{++}\gamma$     & $3.72^{+0.14}_{-0.13}$       & $3.62174\pm0.0001 $(exp)          & $-6.388^{+2.367}_{-3.457}$ & $-0.219^{+0.102}_{-0.223}$ \\
		$\Xi_{cc}^{*+}\to\Xi_{cc}^{+}\gamma$       & $3.72^{+0.14}_{-0.13}$       & $3.61997\pm0.00209$(exp)   & $-1.507^{+0.728}_{-1.369}$ & $0.109^{+0.111}_{-0.058}$  \\
		$\Omega_{cc}^{*+}\to\Omega_{cc}^{+}\gamma$ & $3.82^{+0.13}_{-0.15}$       & $3.7259\pm0.0012$(exp)            & $-1.999^{+0.890}_{-1.232}$ & $0.056^{+0.034}_{-0.027}$  \\
		$\Xi_{bb}^{*0}\to\Xi_{bb}^{0}\gamma$       & $10.30^{+0.14}_{-0.18} $       & $10.27^{+0.15}_{-0.19}$           & $-4.490^{+1.554}_{-1.923}$ & $-0.157^{+0.075}_{-0.082}$ \\
		$\Xi_{bb}^{*-}\to\Xi_{bb}^{-}\gamma$       & $10.30^{+0.14}_{-0.18} $       & $10.27^{+0.15}_{-0.19}         $             & $4.883^{+1.922}_{-1.538}$  & $0.079^{+0.041}_{-0.038}$  \\
		$\Omega_{bb}^{*-}\to\Omega_{bb}^{-}\gamma$ & $10.34^{+0.15}_{-0.18} $       & $10.32^{+0.15}_{-0.19}       $             & $3.880^{+1.484}_{-1.183}$  & $0.041^{+0.022}_{-0.020}$  \\
		\hline\hline
	\end{tabular}
\end{center}
\end{table*}

\begin{table*}[hbtp]
	\begin{center}
   \renewcommand{\arraystretch}{1.6}
	\caption{The results of multipole form factors $G_{M1}(0)$ and $G_{E2}(0)$, transition magnetic moments $\mu$ (in units of $\mu_N$), and radiative widths $\Gamma$ (in keV). We use the magnetic-dipole form factors $G^l_{M1}(0)$ and $G^h_{M1}(0)$ to reflect the contributions through the light-quark light-cone distribution amplitudes and direct emission from heavy-quark lines separately }
	\label{tab:em_result}
	\begin{tabular}{lcccccc}
		\hline\hline
		Channel                     &  $G^l_{M1}(0)$&$G^h_{M1}(0)$             & $G_{M1}(0)$ & $G_{E2}(0)$ & $\mu$ [$\mu_N$] & $\Gamma$ [keV]        \\
		\hline
		$\Xi_{cc}^{*++}\to\Xi_{cc}^{++}\gamma$  &$-2.153^{+0.779}_{-0.996}$&$-2.076^{+0.818}_{-1.342}$   & $-4.229^{+1.567}_{-2.289}$    & $-0.029^{+0.011}_{-0.017}$    & $-1.11^{+0.41}_{-0.60}$        & $5.25^{+8.57}_{-3.62}$               \\
		$\Xi_{cc}^{*+}\to\Xi_{cc}^{+}\gamma$    &$1.074^{+0.497}_{-0.389}$&$ -2.072^{+1.566}_{-1.566} $ & $-0.998^{+0.482}_{-0.906}$    & $-0.007^{+0.003}_{-0.006}$    & $-0.26^{+0.13}_{-0.24}$        & $0.31^{+0.84}_{-0.25}$               \\
		$\Omega_{cc}^{*+}\to\Omega_{cc}^{+}\gamma$ &$0.762^{+0.345}_{-0.270}$&$-2.087^{+0.821}_{-1.126}$ & $-1.324^{+0.590}_{-0.816}$    & $-0.008^{+0.004}_{-0.005}$    & $-0.34^{+0.15}_{-0.21}$        & $0.41^{+0.67}_{-0.30}$               \\
		$\Xi_{bb}^{*0}\to\Xi_{bb}^{0}\gamma$   &$-4.162^{+1.356}_{-1.691}$&$1.170^{+0.456}_{-0.371}$    & $-2.991^{+1.035}_{-1.281}$    & $-0.002^{+0.001}_{-0.001}$    & $-0.27^{+0.09}_{-0.12}$        & $0.01^{+0.01}_{-0.01}$ \\
		$\Xi_{bb}^{*-}\to\Xi_{bb}^{-}\gamma$  &$2.082^{+0.846}_{-0.679}$&$1.170^{+0.456}_{-0.371}$     & $3.253^{+1.280}_{-1.025}$     & $0.002^{+0.001}_{-0.001}$     & $0.30^{+0.12}_{-0.09}$         & $0.01^{+0.02}_{-0.01}$               \\
		$\Omega_{bb}^{*-}\to\Omega_{bb}^{-}\gamma$ &$1.436^{+0.571}_{-0.460}$&$1.149^{+0.438}_{-0.352}$ & $2.585^{+0.989}_{-0.788}$     & $0.001^{+0.001}_{-0.001}$     & $0.24^{+0.09}_{-0.07}$         & $0.002^{+0.003}_{-0.002}$ \\
		\hline\hline
	\end{tabular}
\end{center}
\end{table*}

\subsection{Comparison with other approaches}
\label{sec:comparison}

Table~\ref{tab:all_comparison} compares the transition magnetic moments and radiative widths for the six channels considered here. The lattice-QCD results are taken from Bahtiyar \etal~\cite{Bahtiyar:2018vub}. Both the heavy-baryon chiral perturbation theory (HBChPT) results and the nonrelativistic quark-model (NRQM) estimates are taken from Li \etal~\cite{Li:2017pxa}; the NRQM column contains the quark-model results reported in that study. The relativistic constituent three-quark-model widths are taken from Branz \etal~\cite{Branz:2010pq}. Magnetic moments are expressed in units of $\mu_N$ and widths in keV. A dash indicates that the corresponding quantity was not reported in the cited work.

\begin{table*}[htbp]
	\centering
	\caption{Comparison of transition magnetic moments and radiative widths across theoretical approaches. Each entry gives $(\mu/\mu_N,\Gamma/{\rm keV})$. For lattice QCD, only the radiative widths directly reported in
Ref.~\cite{Bahtiyar:2018vub} are listed. Transition magnetic moments
are not quoted here, since their comparison requires an explicit
matching of the form-factor and magnetic-moment conventions.}
	\label{tab:all_comparison}
	\begin{tabular}{lccccc}
		\hline\hline
		Channel                                    & This work                     & Lattice QCD~\cite{Bahtiyar:2018vub} & HBChPT~\cite{Li:2017pxa} & NRQM~\cite{Li:2017pxa} & Branz \etal~\cite{Branz:2010pq} \\
		\hline
		$\Xi_{cc}^{*++}\to\Xi_{cc}^{++}\gamma$     & $(-1.11^{+0.41}_{-0.60},        5.25^{+8.57}_{-3.62})$              & $(\text{--},0.0518(56))$                     & $(-2.35,22.0)$           & $(-1.40,\text{--})$    & $(\text{--},23.46\pm3.33)$             \\
		$\Xi_{cc}^{*+}\to\Xi_{cc}^{+}\gamma$       & $(-0.26^{+0.13}_{-0.24},0.31^{+0.84}_{-0.25})$              & $(\text{--},0.0648(38))$                     & $(+1.55,9.57)$           & $(+1.23,\text{--})$    & $(\text{--},28.79\pm2.51)$             \\
		$\Omega_{cc}^{*+}\to\Omega_{cc}^{+}\gamma$ & $(-0.34^{+0.15}_{-0.21},0.41^{+0.67}_{-0.30})$              & $(\text{--},0.0565(4))$                     & $(+1.54,9.45)$           & $(+0.90,\text{--})$    & $(\text{--},2.11\pm0.11)$             \\
		\hline
		$\Xi_{bb}^{*0}\to\Xi_{bb}^{0}\gamma$       & $(-0.27^{+0.09}_{-0.12},0.01^{+0.01}_{-0.01})$  & $\text{--}$                         & $(-2.77,31.1)$              & $(-1.82,\text{--})$    & $(\text{--},0.31\pm0.06)$             \\
		$\Xi_{bb}^{*-}\to\Xi_{bb}^{-}\gamma$       & $(0.30^{+0.12}_{-0.09},0.01^{+0.02}_{-0.01})$              & $\text{--}$                         & $(+1.13,5.17)$              & $(+0.81,\text{--})$    & $(\text{--},0.0587\pm0.0142)$             \\
		$\Omega_{bb}^{*-}\to\Omega_{bb}^{-}\gamma$ & $(0.24^{+0.09}_{-0.07},0.002^{+0.003}_{-0.002})$ & $\text{--}$                         & $(+1.12,5.08)$              & $(+0.48,\text{--})$    & $(\text{--},0.0226\pm0.0045)$             \\
		\hline\hline
	\end{tabular}
\end{table*}

For the six ground-state transitions $B^*(3/2^+)\to B(1/2^+)\gamma$,  the central values of our
radiative widths are smaller than both the HBChPT predictions
and the relativistic constituent quark-model results of
Branz et al. In contrast, for the three doubly charmed channels,
our central values are larger than the lattice-QCD results. These differences may arise from the baryon masses, hyperfine splittings, form-factor normalizations, and treatments of intermediate hadronic states used in the respective approaches. In particular, the width depends on the kinematic factor $(M_{B^*}^2-M_B^2)^3$. Even a moderate difference in the predicted hyperfine splitting can therefore lead to a much larger difference in $\Gamma$.

The sign of a transition magnetic moment changes when one external baryon state is rephased. Comparisons of absolute signs therefore require consistent interpolating-current and electromagnetic-current conventions. Within a given convention, the relative signs of the light- and heavy-quark contributions provide information about their interference and the underlying dynamics.

\section{ Summary and discussions}
\label{sec:include}

We evaluate photon emission through the light-quark light-cone distribution amplitudes and direct emission from heavy-quark lines separately. The corresponding spectral densities are denoted by $\rho_\alpha^{l}(s)$ and $\rho_\alpha^{h}(s)$. Combining the two contributions gives the full results for $G_{M_1}(0)$, $G_{E_2}(0)$, and $\mu$ listed in Table~\ref{tab:em_result}. 

For the $\Xi_{cc}^{*+}\to\Xi_{cc}^{+}\gamma$ and
$\Omega_{cc}^{*+}\to\Omega_{cc}^{+}\gamma$ transitions, our
$G_{M1}(0)$ values have signs opposite to those reported in the
lattice-QCD calculation~\cite{Bahtiyar:2018vub}. However, this
difference should be interpreted only after matching the
external-state phase and form-factor conventions used in the
two approaches. In our decomposition, the heavy-quark contribution has the opposite sign to the light-quark contribution and a larger magnitude in both channels, so it determines the sign of the total transition moment. The lattice calculation exhibits a different pattern: in its adopted convention, the charm-sector contribution has the opposite sign to ours and partially cancels the light-sector contribution, leaving a positive magnetic-dipole form factor $G_{M1}(0)$.

The lattice calculation also allows a comparison of the individual quark-sector contributions. At $Q^2=0$, the charm-sector contributions to $G_{M1}$ are approximately $G_{M1}^{c}(0)=0.528(112)$ for the $\Xi_{cc}^{*+}$ transitions and $G_{M1}^{c}(0)=0.571(24)$ for the $\Omega_{cc}^{*+}$ transition in Ref.~\cite{Bahtiyar:2018vub}. The corresponding light-sector contributions are approximately $-1.763(64)$ and $-1.504(32)$, respectively. Thus, the heavy-quark contribution is non-negligible, because it is of the same order of magnitude as the light-sector contribution and substantially affects the total form factor. More generally, interference between the two sectors can make the sign of the transition moment sensitive to their relative magnitudes.

For simplicity, we use a single Borel transformation and use the approximations $p^2\simeq k^2$. A double Borel transformation in $p^2$ and $k^2$ would treat the initial and final baryon poles independently and is formally more appropriate for transitions between distinct states. The single-Borel prescription may therefore introduce additional uncertainties in pole residues, continuum subtraction, and extracted form factors $G_1(0)$ and $G_2(0)$. These uncertainties are particularly relevant to the transition moment and radiative widths, since $\Gamma$ depends quadratically on the magnetic form factor and cubically on mass splitting.

Additional uncertainties arise from baryon masses, continuum thresholds, Borel windows, condensate parameters, and distribution amplitudes. The $E2$ form factors are much smaller than the $M1$ form factors in all six channels. Consequently, uncertainties in the widths are primarily associated with $G_{M1}(0)$, including the relative contributions of light- and heavy-quark, and with the baryon masses.

In conclusion, we systematically investigate the masses of the low-lying doubly heavy baryons and their $B^*(3/2^+)\to B(1/2^+)\gamma$ radiative transitions within QCD sum rules and light-cone sum rules. The extracted masses are compatible with the observed states $\Xi_{cc}(1/2^+)$ and $\Omega_{cc}(1/2^+)$ and predict hyperfine splittings of about $108$ MeV and $99$ MeV for the $\Xi_{cc}$ and $\Omega_{cc}$ systems, respectively. For radiative transitions, the contribution of electric-quadrupole form factor is small, and the widths are dominated by the magnetic-dipole form factor. Comparisons with lattice QCD, HBChPT, and nonrelativistic and relativistic quark models reveal substantial differences among the predictions. Although the light-quark contribution generally dominates, the heavy-quark contribution cannot be neglected. Its relative sign is particularly important to understand the differences between our results and other studies. A future analysis using a double Borel transformation, together with a detailed separation of the light- and heavy-quark photon-emission contributions, would help establish more robust predictions for the transition moments and decay widths.

\begin{acknowledgements}
This project is supported by the National Natural Science Foundation of China under Grant No.~12675096, the Jiangsu Provincial Double-Innovation Program under Grant No.~JSSCRC2021488, and the Fundamental Research Funds for the Central Universities.
\end{acknowledgements}

\appendix

\section{The spectral densities for mass sum rules and light-cone sum rules }
\label{app:ope}

In this appendix, we collect the spectral densities $\rho_{\eta_1^\prime \eta_1^\prime}$, $\rho_{\eta_1\eta_1}$, $\rho_{\eta_1^\prime \eta_1}$ for mass sum rules of $\Omega_{cc}(1/2^+)$.  We also express the spectral densities $\rho_{1}^{l}(s)$, $\rho_{1}^{h}(s)$, $\rho_{2}^{l}(s)$, and $\rho_{2}^{h}(s)$ for radiative transition $\Omega_{cc}^*(3/2^+)\to \Omega_{cc}(1/2^+)\gamma$ within light-cone sum rules. The OPE includes the perturbative contribution, the quark condensate, the mixed quark-gluon condensate, and the gluon condensate corrections.

\begin{widetext}
\begin{eqnarray}
\nonumber \rho_{\eta^\prime_1\eta^\prime_1}(s,M_B^2)&=&\int_{\alpha_{min}}^{\alpha_{\max}}d\alpha\int_{\beta_{min}}^{\beta_{max}}d\beta ({9 F(\alpha,\beta)^2 m_c\over 128 \pi^4\alpha^2\beta}-{15 F(\alpha,\beta)m_c^2 m_s\over 128\pi^4\alpha\beta})+\int_{\alpha_{min}}^{\alpha_{\max}}d\alpha ({3 H(\alpha)^2 m_s\over 256\pi^4(1-\alpha)\alpha}+{9 H(\alpha)m_c m_s^2\over 64\pi^4\alpha})
\\ \nonumber &&+\int_{\alpha_{min}}^{\alpha_{\max}}d\alpha ({3 H(\alpha)\over 32 \pi^2}-{3 m_c^2\over 16\pi^2}+{3m_c m_s (1-\alpha)\over 16\pi^2})\langle\bar s s\rangle+\int_0^1 d\alpha {3 m_c^3 m_s\over 32\pi^2\alpha}\langle\bar s s\rangle\delta(s-{m_c^2\over(1-\alpha)\alpha}) 
\\ \nonumber &&+\int_{\alpha_{min}}^{\alpha_{\max}}d\alpha\int_{\beta_{min}}^{\beta_{max}}d\beta({3 m_c\beta \over 512\pi^4\alpha^2}-{m_c\over 1024\pi^4\beta})\langle g_s^2 G G\rangle
\\ \nonumber 
&&-\int_0^1 d\alpha \int_0^{1-\alpha}d\beta ({m_c^3\beta\over 512\pi^4\alpha^3}+{m_c^3\over 512\pi^4\alpha^2})\langle g_s^2 G G\rangle\delta(s-{m_c^2(\alpha+\beta)\over \alpha\beta})
\\ \nonumber &&+\int_{\alpha_{min}}^{\alpha_{\max}}d\alpha({3 m_c\over 1024\pi^4}+{m_c(1-\alpha)\over 1024\pi^4\alpha})\langle g_s^2 G G\rangle
+\int_{\alpha_{min}}^{\alpha_{\max}}d\alpha({3(1-\alpha)\alpha\over 64\pi^2}-{1-\alpha\over 64\pi^2})\langle g_s\bar s \sigma G s\rangle
\\ \nonumber &&+\int_0^1 d\alpha ({9 m_c^2\over 128 \pi^2}-{7m_c^2\over 128\pi^2\alpha})\langle g_s \bar s \sigma G s\rangle\delta(s-{m_c^2\over (1-\alpha)\alpha})
+\int_0^1 d\alpha {1\over 1536 \pi^2}\langle \bar s s\rangle \langle g_s^2 G G\rangle \delta(s-{m_c^2\over (1-\alpha)\alpha})
\\ \nonumber &&
+{1\over M_B^2}\times \left\{\int_0^1 d\alpha {6m_c^4\over 128\pi^2(1-\alpha)\alpha}\langle g_s\bar s\sigma G s\rangle \delta(s-{m_c^2\over (1-\alpha)\alpha})\right.
\\ \nonumber &&+\left.\int_0^1 d\alpha (-{5 m_c^2\over 384 \pi^2(1-\alpha)^2}+{m_c^2\over 512 \pi^2(1-\alpha)\alpha})\langle\bar s s\rangle \langle g_s^2 GG\rangle \delta(s-{m_c^2\over (1-\alpha)\alpha})\right\}
\\ \nonumber &&+{1\over M_B^4}\times\left\{\int_0^1 d\alpha {m_c^4\over 192\pi^2(1-\alpha)^3}\langle\bar s s\rangle\langle g_s^2 GG\rangle\delta(s-{m_c^2\over (1-\alpha)\alpha})\right.
\\ \nonumber && \left.+\int_0^1 d\alpha (-{5m_c^2\over 1536\pi^2(1-\alpha)^2}+{m_c^2\over 3072\pi^2(1-\alpha)\alpha})\langle g_s^2 GG\rangle\langle g_s\bar s\sigma G s\rangle\delta(s-{m_c^2\over (1-\alpha)\alpha})\right\}
\\ \nonumber &&+{1\over M_B^6}\times \int_0^1 d\alpha({m_c^4\over 384\pi^2(1-\alpha)^3}+{5 m_c^4\over 1536\pi^2(1-\alpha)^3\alpha}-{m_c^4\over 2048\pi^2(1-\alpha)^2\alpha^2})\langle g_s^2 GG\rangle\langle g_s\bar s\sigma G s\rangle\delta(s-{m_c^2\over (1-\alpha)\alpha}) 
\\  &&+{1\over M_B^8}\times\int_0^1 d\alpha  {-m_c^6\over 768\pi^2(1-\alpha)^4\alpha}\langle g_s^2 G G\rangle\langle g_s\bar s\sigma G s\rangle\delta(s-{m_c^2\over (1-\alpha)\alpha}) \, .
\\ \nonumber \rho_{\eta_1\eta_1}(s,M_B^2)&=&\int_{\alpha_{min}}^{\alpha_{\max}}d\alpha\int_{\beta_{min}}^{\beta_{max}}d\beta {-3 F(\alpha,\beta)m_c^2 m_s\over 4\pi^4\alpha\beta}+\int_{\alpha_{min}}^{\alpha_{\max}}d\alpha {3 H(\alpha)^2 m_s\over 16\pi^4(1-\alpha)\alpha}
+\int_{\alpha_{min}}^{\alpha_{\max}}d\alpha ({3 H(\alpha)\over 2\pi^2}-{3 m_c^2\over 2\pi^2})\langle \bar s s\rangle
\\ \nonumber &&+\int_{\alpha_{min}}^{\alpha_{\max}}d\alpha ({3(1-\alpha)\alpha\over 4\pi^2}-{1-\alpha\over 4\pi^2})\langle g_s \bar s\sigma G s\rangle+\int_0^1 d\alpha ({3 m_c^2\over 4\pi^2}-{3m_c^2\over 8\pi^2\alpha})\langle g_s \bar s\sigma G s\rangle\delta(s-{m_c^2\over (1-\alpha)\alpha})
\\ \nonumber &&+\int_0^1 d\alpha {1\over 96\pi^2}\langle\bar s s\rangle\langle g_s^2 G G\rangle\delta(s-{m_c^2\over(1-\alpha)\alpha})+{1\over M_B^2}\times \left\{{3 m_c^4\over 8\pi^2(1-\alpha)\alpha}\langle g_s\bar s \sigma G s\rangle\delta(s-{m_c^2\over(1-\alpha)\alpha})\right.
\\ \nonumber &&\left.+\int_0^1 d\alpha ({-m_c^2\over 12\pi^2(1-\alpha)^2}+{m_c^2\over 96\pi^2(1-\alpha)\alpha})\langle\bar s s\rangle\langle g_s^2 G G \rangle\delta(s-{m_c^2\over(1-\alpha)\alpha})\right\}
\\ \nonumber &&+{1\over M_B^4}\times\left\{\int_0^1 d\alpha {m_c^4\over 24\pi^2(1-\alpha)^3}\langle\bar s s\rangle\langle g_s^2 G G\rangle\delta(s-{m_c^2\over (1-\alpha)\alpha})\right.
\\ \nonumber &&+\left.\int_0^1 d\alpha {-m_c^2\over 48\pi^2(1-\alpha)^2}\langle g_s^2 G G\rangle\langle g_s \bar s\sigma G s\rangle\delta(s-{m_c^2\over (1-\alpha)\alpha})\right\}
\\ \nonumber &&+{1\over M_B^6}\times \int_0^1 d\alpha ({m_c^4\over 48\pi^2(1-\alpha)^3}+{m_c^4\over 48 \pi^2(1-\alpha)^3\alpha}-{m_c^4\over 384\pi^2(1-\alpha)^2\alpha^2})\langle g_s^2 G G \rangle\langle g_s \bar s \sigma G s\rangle\delta(s-{m_c^2\over(1-\alpha)\alpha})
\\  &&+{1\over M_B^8}\times \int_0^1 d\alpha {-m_c^6\over 96 \pi^2(1-\alpha)^4\alpha}\langle g_s^2 G G \rangle\langle g_s \bar s \sigma G s\rangle\delta(s-{m_c^2\over(1-\alpha)\alpha})\, .
\\ \nonumber \rho_{\eta^\prime_1\eta_1}(s,M_B^2)&=&\int_{\alpha_{min}}^{\alpha_{\max}}d\alpha\int_{\beta_{min}}^{\beta_{max}}d\beta ({-9 F(\alpha,\beta)m_c\over 64\pi^4\alpha^2\beta}+{3 F(\alpha,\beta)m_c^2 m_s\over 16 \pi^4\alpha\beta})-\int_{\alpha_{min}}^{\alpha_{\max}}d\alpha({3 H(\alpha)^2 m_s\over 64\pi^4(1-\alpha)\alpha}+{9H(\alpha)m_c m_s^2\over 32\pi^4\alpha})
\\ \nonumber &&+\int_{\alpha_{min}}^{\alpha_{\max}}d\alpha ({-3H(\alpha)\over 8\pi^2}+{3m_c^2\over 8\pi^2}-{3m_c m_s(1-\alpha)\over 8\pi^2})\langle\bar s s\rangle-\int_0^1 d\alpha {3m_c^3 m_s\over 16\pi^2\alpha}\langle\bar s s\rangle\delta(s-{m_c^2\over(1-\alpha)\alpha})
\\ \nonumber &&+\int_{\alpha_{min}}^{\alpha_{\max}}d\alpha\int_{\beta_{min}}^{\beta_{max}}d\beta({m_c\over 512\pi^4\beta}-{3m_c\beta\over 256\pi^4\alpha^2})\langle g_s^2 G G\rangle
\\ \nonumber &&+\int_0^1 d\alpha\int_0^{1-\alpha}d\beta ({m_c^3\beta\over 256\pi^4\alpha^3}+{m_c^3\over 256\pi^4\alpha^2})\langle g_s^2 G G\rangle\delta(s-{m_c^2(\alpha+\beta)\over\alpha\beta})
\\ \nonumber &&-\int_{\alpha_{min}}^{\alpha_{\max}}d\alpha ({3m_c\over 512\pi^4}+{m_c(1-\alpha)\over 512\pi^4\alpha})\langle g_s^2 GG \rangle-\int_{\alpha_{min}}^{\alpha_{\max}}d\alpha({3(1-\alpha)\alpha\over 16\pi^2}-{1-\alpha\over 16\pi^2})\langle g_s\bar s\sigma G s\rangle
\\ \nonumber &&-\int_0^1({3m_c^2\over 16\pi^2}-{3m_c^2\over 32\pi^2\alpha})\langle g_s \bar s\sigma G s\rangle \delta(s-{m_c^2\over(1-\alpha)\alpha})
-\int_0^1 d\alpha {1\over 384\pi^2}\langle \bar s s\rangle\langle g_s^2 G G\rangle\delta(s-{m_c^2\over(1-\alpha)\alpha})
\\\nonumber &&+{1\over M_B^2}\times\left\{\int_0^1 d\alpha {-3 m_c^4\over 32\pi^2(1\alpha)\alpha}\langle g_s\bar s\sigma G s\rangle\delta(s-{m_c^2\over(1-\alpha)\alpha})\right.
\\ \nonumber &&\left.+\int_0^1 d\alpha ({m_c^2\over 48\pi^2(1-\alpha)^2}-{m_c^2\over 384\pi^2(1-\alpha)\alpha})\langle \bar s s\rangle\langle g_s^2 G G\rangle\delta(s-{m_c^2\over(1-\alpha)\alpha})\right\}
\\ \nonumber &&+{1\over M_B^4}\times\left\{\int_0^1 d\alpha {-m_c^4\over 96\pi^2(1-\alpha)^3}\langle \bar s s\rangle\langle g_s^2 G G\rangle\delta(s-{m_c^2\over(1-\alpha)\alpha})\right.
\\ \nonumber &&+\left.\int_0^1 d\alpha{m_c^2\over 192\pi^2(1-\alpha)^2}\langle g_s^2 G G \rangle\langle g_s \bar s\sigma G s\rangle\delta(s-{m_c^2\over(1-\alpha)\alpha}) \right\}
\\ \nonumber &&+{1\over M_B^6}\times \int_0^1 d \alpha ({-m_c^4\over 192\pi^2(1-\alpha)^3}-{m_c^4\over 192\pi^2(1-\alpha)^3\alpha}+{m_c^4\over 1536\pi^2(1-\alpha)^2\alpha^2})\langle g_s^2 G G \rangle\langle g_s \bar s\sigma G s\rangle\delta(s-{m_c^2\over(1-\alpha)\alpha})
\\  &&+{1\over M_B^8}\times\int_0^1 d \alpha {m_c^6\over 384\pi^2(1-\alpha)^4\alpha}\langle g_s^2 G G \rangle\langle g_s \bar s\sigma G s\rangle\delta(s-{m_c^2\over(1-\alpha)\alpha})\, .
\\ \nonumber \rho^l_1(s,M_B)&=&e_s\times\left\{\int_{\alpha_{min}}^{\alpha_{\max}}d\alpha\int_0^1 d u\Big({-3m_c\chi\cos\theta_2^\prime t_1\over8\pi^2}\langle\bar s s\rangle\phi_{\gamma}(u)-{5 m_c\chi\sin\theta_2^\prime t_2\over4\pi^2}\langle\bar s s\rangle\phi_\gamma(u)\right.
\\ \nonumber &&-\left.{5m_c\chi\cos\theta_2^\prime t_2\over 8\pi^2}\langle\bar s s\rangle\phi_\gamma(u)+{3 f_{3_\gamma}\sin\theta_2^\prime t_1\alpha\over 8\pi^2}\psi^{(a)}(u)+{3 f_{3_\gamma}\cos\theta_2^\prime t_1\alpha\over 16\pi^2}\psi^{(a)}(u)\right.
\\ \nonumber &&+\left.{3 m_c\chi \cos\theta_2^\prime t_1\alpha\over 8\pi^2}\langle\bar s s\rangle\phi_\gamma(u)+{5 m_c \chi\sin\theta_2^\prime t_2\alpha\over4\pi^2}\langle \bar s s\rangle\phi_\gamma(u)
+{5 m_c\chi\cos\theta_2^\prime t_2\alpha\over8\pi^2}\langle\bar s s\rangle\phi_\gamma(u)\right.
\\ \nonumber &&-\left.{3 f_{3_\gamma}\sin\theta_2^\prime t_1 \alpha^2\over 8\pi^2}\psi^{(a)}(u)-{3 f_{3_\gamma}\cos\theta_2^\prime t_1 \alpha^2\over 16\pi^2}\psi^{(a)}(u)\Big)+\int_0^1 d\alpha\int_0^1 d u\Big({5 m_c^2 f_{3_\gamma}\sin\theta_2^\prime t_1\over 16\pi^2}\psi^{(a)}(u)\right.
\\ \nonumber &&
+{5 m_c^2 f_{3_\gamma}\cos\theta_2^\prime t_1\over 32\pi^2}\psi^{(a)}(u)+{3 m_c^2 f_{3_\gamma}\cos\theta_2^\prime t_2\over 8\pi^2}\psi^{(a)}(u)\Big)\delta(s-{m_c^2\over(1-\alpha)\alpha})
\\ \nonumber &&+\left.\int_0^1 d\alpha\int \mathcal{D}\underline{\alpha}\Big({m_c\cos\theta_2^\prime t_1\over 8\pi^2}\langle\bar s s\rangle S(\underline{\alpha})-{m_c\cos\theta_2^\prime t_1\over 16\pi^2}\langle\bar s s\rangle \widetilde{S}(\underline{\alpha})+{m_c\cos\theta_2^\prime t_1\over 8\pi^2}\langle \bar s s\rangle T_1(\underline{\alpha})\right.
\\ \nonumber &&-\left.{m_c\cos\theta_2^\prime t_1\over8\pi^2}\langle\bar s s\rangle T_2(\underline{\alpha})
+{m_c\cos\theta_2^\prime t_1\over 16\pi^2}\langle \bar s s\rangle T_3(\underline{\alpha})-{m_c\cos\theta_2^\prime t_1\over 16\pi^2}\langle \bar s s\rangle T_4(\underline{\alpha})+{m_c\sin\theta_2^\prime t_2\over 4\pi^2}\langle \bar s s\rangle S(\underline{\alpha})\right.
\\ \nonumber &&+\left.{m_c\sin\theta_2^\prime t_2\over 8\pi^2}\langle \bar s s\rangle \widetilde{S}(\underline{\alpha})
+{m_c\sin\theta_2^\prime t_2\over 4\pi^2}\langle \bar s s\rangle T_1(\underline{\alpha})-{m_c\sin\theta_2^\prime t_2\over 4\pi^2}\langle \bar s s\rangle T_2(\underline{\alpha})-{ 3m_c\sin\theta_2^\prime t_2\over 8\pi^2}\langle \bar s s\rangle T_3(\underline{\alpha})\right.
\\ \nonumber &&+\left.{3 m_c\sin\theta_2^\prime t_2\over 8\pi^2}\langle \bar s s\rangle T_4(\underline{\alpha})
+{m_c\cos\theta_2^\prime t_2\over 8\pi^2}\langle \bar s s\rangle S(\underline{\alpha})+{m_c\cos\theta_2^\prime t_2\over 16\pi^2}\langle \bar s s\rangle \widetilde{S}(\underline{\alpha})+{m_c\cos\theta_2^\prime t_2\over 8\pi^2}\langle \bar s s\rangle T_1(\underline{\alpha})\right.
\\ \nonumber &&-\left.{m_c\cos\theta_2^\prime t_2\over 8\pi^2}\langle \bar s s\rangle T_2(\underline{\alpha})
-{3m_c\cos\theta_2^\prime t_2\over 16\pi^2}\langle \bar s s\rangle T_3(\underline{\alpha})+{3m_c\cos\theta_2^\prime t_2\over 16\pi^2}\langle \bar s s\rangle T_4(\underline{\alpha})+{m_c\cos\theta_2^\prime t_1\over 16\pi^2\alpha}\langle \bar s s\rangle S(\underline{\alpha})\right.
\\ \nonumber &&+\left.{m_c\cos\theta_2^\prime t_1\over 16\pi^2\alpha}\langle \bar s s\rangle \widetilde{S}(\underline{\alpha})
+{m_c\cos\theta_2^\prime t_1\over 16\pi^2\alpha}\langle \bar s s\rangle T_1(\underline{\alpha})-{m_c\cos\theta_2^\prime t_1\over 16\pi^2\alpha}\langle \bar s s\rangle T_2(\underline{\alpha})-{m_c\cos\theta_2^\prime t_1\over 16\pi^2\alpha}\langle \bar s s\rangle T_3(\underline{\alpha})\right.
\\ \nonumber &&+\left.{m_c\cos\theta_2^\prime t_1\over 16\pi^2\alpha}\langle \bar s s\rangle T_4(\underline{\alpha})
+{m_c\sin\theta_2^\prime t_2\over 8\pi^2\alpha}\langle \bar s s\rangle S(\underline{\alpha})-{m_c\sin\theta_2^\prime t_2\over 8\pi^2\alpha}\langle \bar s s\rangle \widetilde{S}(\underline{\alpha})+{3m_c\sin\theta_2^\prime t_2\over 8\pi^2\alpha}\langle \bar s s\rangle T_1(\underline{\alpha})\right.
\\ \nonumber &&-\left.{3m_c\sin\theta_2^\prime t_2\over 8\pi^2\alpha}\langle \bar s s\rangle T_2(\underline{\alpha})
+{3m_c\sin\theta_2^\prime t_2\over 8\pi^2\alpha}\langle \bar s s\rangle T_3(\underline{\alpha})-{3m_c\sin\theta_2^\prime t_2\over 8\pi^2\alpha}\langle \bar s s\rangle T_4(\underline{\alpha})+{m_c\cos\theta_2^\prime t_2\over 16\pi^2\alpha}\langle \bar s s\rangle S(\underline{\alpha})\right.
\\ \nonumber &&-\left.{m_c\cos\theta_2^\prime t_2\over 16\pi^2\alpha}\langle \bar s s\rangle \widetilde{S}(\underline{\alpha})
+{3 m_c\cos\theta_2^\prime t_2\over 16\pi^2\alpha}\langle \bar s s\rangle T_1(\underline{\alpha})-{3 m_c\cos\theta_2^\prime t_2\over 16\pi^2\alpha}\langle \bar s s\rangle T_2(\underline{\alpha})+{3 m_c\cos\theta_2^\prime t_2\over 16\pi^2\alpha}\langle \bar s s\rangle T_3(\underline{\alpha})\right.
\\ \nonumber &&-\left.{3 m_c\cos\theta_2^\prime t_2\over 16\pi^2\alpha}\langle \bar s s\rangle T_4(\underline{\alpha})\Big)\delta(s-{m_c^2\over (1-\alpha)\alpha})\right.
\\ \nonumber &&+\left.{1\over M_B^2}\times\int_0^1 d \alpha \int_0^1 du \Big({-f_{3_\gamma}\sin\theta_2^\prime t_1\over 576\pi^2}\langle g_s^2 G G\rangle \psi^{(a)}(u)-{f_{3_\gamma}\cos\theta_2^\prime t_1\over 1152\pi^2}\langle g_s G G\rangle \psi^{(a)}(u)\right.
\\ \nonumber &&-\left.{m_c\chi \cos\theta_2^\prime t_1\over 64\pi^2\alpha^2}\langle\bar s s\rangle\langle g_s^2 G G\rangle\phi_\gamma(u)
-{5 m_c\chi \sin\theta_2^\prime t_2\over 96\pi^2\alpha^2}\langle\bar s s\rangle\langle g_s^2 G G\rangle\phi_\gamma(u)-{5m_c\chi \cos\theta_2^\prime t_2\over 192\pi^2\alpha^2}\langle\bar s s\rangle\langle g_s^2 G G\rangle\phi_\gamma(u)\right.
\\ \nonumber &&+\left.{7m_c\chi \cos\theta_2^\prime t_1\over 384\pi^2\alpha}\langle\bar s s\rangle\langle g_s^2 G G\rangle\phi_\gamma(u)
+{35m_c\chi \sin\theta_2^\prime t_2\over 576\pi^2\alpha}\langle\bar s s\rangle\langle g_s^2 G G\rangle\phi_\gamma(u)\right.
\\ \nonumber &&+\left.{35m_c\chi \cos\theta_2^\prime t_2\over 1152\pi^2\alpha}\langle\bar s s\rangle\langle g_s^2 G G\rangle\phi_\gamma(u)\Big)\delta(s-{m_c^2\over(1-\alpha)\alpha})\right.
\\ \nonumber &&+\left.{1\over M_B^4}\times \int_0^1 d\alpha\int_0^1 du\Big({m_c^2 f_{3_\gamma}\sin\theta_2^\prime t_1\over 64\pi^2(1-\alpha)^2}\langle g_s^2 G G\rangle\psi^{(a)}(u)+{m_c^2 f_{3_\gamma}\cos\theta_2^\prime t_1\over 128\pi^2(1-\alpha)^2}\langle g_s^2 G G\rangle\psi^{(a)}(u)\right.
\\ \nonumber &&+\left.{m_c^3 \chi\cos\theta_2^\prime t_1\over 192\pi^2(1-\alpha)^2}\langle \bar s s\rangle\phi_\gamma(u)
+{5m_c^3\chi\sin\theta_2^\prime t_2\over 288\pi^2(1-\alpha)^2}\langle\bar s s\rangle \langle g_s^2 G G\rangle\phi_\gamma(u)+{m_c^2 f_{3_\gamma}\cos\theta_2^\prime t_2\over 32\pi^2(1-\alpha)^2}\langle g_s^2 G G\rangle\psi^{(a)}(u)\right.
\\ \nonumber &&+\left.{5m_c^3 \chi\cos\theta_2^\prime t_2\over 576\pi^2(1-\alpha)^2}\langle \bar s s\rangle\phi_\gamma(u)
+{m_c^3\chi\cos\theta_2^\prime t_1\over 192\pi^2\alpha^3}\langle \bar s s\rangle\phi_\gamma(u)+{5m_c^3 \chi\sin\theta_2^\prime t_2\over 288\pi^2\alpha^3}\langle\bar s s\rangle\phi_\gamma(u)\right.
\\ \nonumber &&+\left.{5m_c^3\chi\cos\theta_2^\prime t_2\over 576\pi^2\alpha^3}\langle \bar s s\rangle\phi_\gamma(u)
-{m_c^3\chi\cos\theta_2^\prime t_1\over 192\pi^2\alpha^2}\langle \bar s s\rangle\phi_\gamma(u)-{5m_c^3\chi\sin\theta_2^\prime t_2\over 288\pi^2\alpha^2}\langle \bar s s\rangle\phi_\gamma(u)\right.
\\ \nonumber &&-\left.{5m_c^3\chi\cos\theta_2^\prime t_2\over 576\pi^2\alpha^2}\langle \bar s s\rangle\phi_\gamma(u)
-{m_c^2 f_{3_\gamma}\sin\theta_2^\prime t_1\over 576\pi^2(1-\alpha)\alpha}\langle g_s^2 G G\rangle\psi^{(a)}(u)-{m_c^2 f_{3_\gamma}\cos\theta_2^\prime t_1\over 1152\pi^2(1-\alpha)\alpha}\langle g_s^2 G G\rangle\psi^{(a)}(u)\right.
\\ \nonumber &&-\left.{m_c^2 f_{3_\gamma}\cos\theta_2^\prime t_2\over 192\pi^2(1-\alpha)\alpha}\langle g_s^2 G G\rangle\psi^{(a)}(u)\Big)\delta(s-{m_c^2\over(1-\alpha)\alpha})\right.
\\ \nonumber&&+\left.{1\over M_B^6}\times\int_0^1 d\alpha\int_0^1 d u\Big({-5m_c^4 f_{3_\gamma}\sin\theta_2^\prime t_1\over 576\pi^2(1-\alpha)^3}\langle g_s^2 G G\rangle\psi^{(a)}(u)-{5m_c^4 f_{3_\gamma}\cos\theta_2^\prime t_1\over 1152\pi^2(1-\alpha)^3}\langle g_s^2 G G\rangle\psi^{(a)}(u)\right.
\\  &&-\left.{m_c^4 f_{3_\gamma}\cos\theta_2^\prime t_2\over 96\pi^2(1-\alpha)^3}\langle g_s^2 G G\rangle\psi^{(a)}(u)\Big)\delta(s-{m_c^2\over(1-\alpha)\alpha})\right\}\, .
\\ \nonumber  \rho^h_1(s,M_B)&=&e_c\times\left\{\int_{\alpha_{min}}^{\alpha_{\max}}d\alpha\int_{\beta_{min}}^{\beta_{max}}d\beta\Big({-3m_c^2\sin\theta_2^\prime t_1\over 16\pi^4}-{3m_c^2\cos\theta_2^\prime t_1\over 32\pi^4}+{9m_c m_s\cos\theta_2^\prime t_1\over 32\pi^4}+{9m_c m_s\sin\theta_2^\prime t_2\over 16\pi^4}\right.
\\ \nonumber &&-\left.{3m_c^2\cos\theta_2^\prime t_2\over 4\pi^4}
+{9m_c m_s\cos\theta_2^\prime t_2\over 32\pi^4}-{3\sin\theta_2^\prime t_1 F(\alpha,\beta)\over 16\pi^4\alpha}+{3m_c^2\sin\theta_2^\prime t_1\over 16\pi^4\alpha}-{3\cos\theta_2^\prime t_1 F(\alpha,\beta)\over 32\pi^4\alpha}\right.
\\ \nonumber &&+\left.{3m_c^2\cos\theta_2^\prime t_1\over 32\pi^4\alpha}+{3m_c^2\cos\theta_2^\prime t_2\over 4\pi^4\alpha}
-{3m_c^2\sin\theta_2^\prime t_1\beta\over 16\pi^4\alpha}-{3m_c^2\cos\theta_2^\prime t_1\beta\over 32\pi^4\alpha}+{3m_c m_s\cos\theta_2^\prime t_1\beta \over 32\pi^4\alpha}\right.
\\ \nonumber &&+\left.{3m_c m_s\sin\theta_2^\prime t_2\beta\over 16\pi^4}-{3m_c^2\cos\theta_2^\prime t_2\beta\over 4\pi^4\alpha}+{3m_c m_s\cos\theta_2^\prime t_2\beta\over 32\pi^4\alpha}\Big)\right.
\\ \nonumber &&+\left.\int_0^1 d\alpha \Big({m_s\sin\theta_2^\prime t_1\over 4\pi^2}-{m_c\cos\theta_2^\prime t_1\over 4\pi^2}+{m_s\cos\theta_2^\prime t_1\over 8\pi^2}-{m_c\sin\theta_2^\prime t_2\over 2\pi^2}-{m_c\cos\theta_2^\prime t_2\over 4\pi^2}-{m_c\cos\theta_2^\prime t_1\over 8\pi^2\alpha}\right.
\\ \nonumber &&-\left.{m_c\sin\theta_2^\prime t_2\over 4\pi^2\alpha}-{m_c\cos\theta_2^\prime t_2\over 8\pi^2\alpha}-{m_s\sin\theta_2^\prime t_1\alpha\over 4\pi^2}-{m_s\cos\theta_2^\prime t_1\alpha\over 8\pi^2}\Big)\langle \bar s s\rangle \delta(s-{m_c^2\over(1-\alpha)\alpha})\right.
\\ \nonumber &&+\left.{1\over M_B^2}\times \int_0^1 d\alpha\Big(\big({m_c\cos\theta_2^\prime t_1\over 16\pi^2}+{m_c\sin\theta_2^\prime t_2\over 8\pi^2}+{m_c\cos\theta_2^\prime t_2\over 16\pi^2}+{m_c\cos\theta_2^\prime t_1\over 32\pi^2\alpha}+{m_c\sin\theta_2^\prime t_2\over 16\pi^2\alpha}\right.
\\ \nonumber &&+\left.{m_c\cos\theta_2^\prime t_2\over 32\pi^2\alpha}\big)\langle g_s\bar s\sigma G s\rangle
+\big({m_c^2 m_s\sin\theta_2^\prime t_1\over 4\pi^2\alpha}+{m_c^2 m_s\cos\theta_2^\prime t_1\over 8\pi^2\alpha}+{m_c^2 m_s\cos\theta_2^\prime t_2\over 2\pi^2\alpha}\big)\langle\bar s s\rangle\Big)\delta(s-{m_c^2\over(1-\alpha)\alpha})\right.
\\ \nonumber &&+\left.{1\over M_B^4}\times\int_0^1 d\alpha\Big({m_c^3\cos\theta_2^\prime t_1\over 32\pi^2\alpha^2}+{m_c^3\sin\theta_2^\prime t_2\over 16\pi^2\alpha^2}+{m_c^3\cos\theta_2^\prime t_2\over 32\pi^2\alpha^2}+{3m_c^3\cos\theta_2^\prime t_1\over 32\pi^2(1-\alpha)\alpha}+{3m_c^3\sin\theta_2^\prime t_2\over 16\pi^2(1-\alpha)\alpha}\right.
\\  &&+\left.{3m_c^3\cos\theta_2^\prime t_2\over 32\pi^2(1-\alpha)\alpha}\Big)\langle g_s\bar s\sigma G s\rangle\delta(s-{m_c^2\over(1-\alpha)\alpha})\right\}\, .
\\ \nonumber \rho_2^l(s,M_B)&=&e_s\times\left\{\int_0^1 d\alpha\int \mathcal{D}\underline{\alpha}\Big({-f_{3_\gamma}\sin\theta_2^\prime t_1\over 32\pi^2}\mathcal{A}(\underline{\alpha})-{f_{3_\gamma}\sin\theta_2^\prime t_1\over 8\pi^2}\mathcal{V}(\underline{\alpha})+{3f_{3_\gamma}\cos\theta_2^\prime t_1\over 256\pi^2}\mathcal{A}(\underline{\alpha})\right.
\\ \nonumber &&-\left.{f_{3_\gamma}\cos\theta_2^\prime t_1\over 32\pi^2}\mathcal{V}(\underline{\alpha})-{f_{3_\gamma}\cos\theta_2^\prime t_2\over 256\pi^2}\mathcal{A}(\underline{\alpha})-{f_{3_\gamma}\cos\theta_2^\prime t_2\over 4\pi^2}\mathcal{V}(\underline{\alpha})+{f_{3_\gamma}\sin\theta_2^\prime t_1\alpha\over 32\pi^2}\mathcal{A}(\underline{\alpha})+{f_{3_\gamma}\sin\theta_2^\prime t_1\alpha\over 8\pi^2}\mathcal{V}(\underline{\alpha})\right.
\\ \nonumber&&-\left.{3f_{3_\gamma}\cos\theta_2^\prime t_1\alpha\over 256\pi^2}\mathcal{A}(\underline{\alpha})+{f_{3_\gamma}\cos\theta_2^\prime t_1\alpha\over 32\pi^2}\mathcal{V}(\underline{\alpha})+{f_{3_\gamma}\cos\theta_2^\prime t_2\alpha\over 256\pi^2}\mathcal{A}(\underline{\alpha})+{f_{3_\gamma}\cos\theta_2^\prime t_2\alpha\over 4\pi^2}\mathcal{V}(\underline{\alpha})\Big)\delta(s-{m_c^2\over(1-\alpha)\alpha})\right.
\\ \nonumber &&+\left.{i\over M_B^2}\times\int_0^1 d\alpha\int \mathcal{D}\underline{\alpha}\Big({-m_c \cos\theta_2^\prime t_1\over 16\pi^2}\widetilde{T}_1(\underline{\alpha})+{7m_c \cos\theta_2^\prime t_1\over 16\pi^2}\widetilde{T}_2(\underline{\alpha})+{7m_c \cos\theta_2^\prime t_1\over 16\pi^2}\widetilde{T}_3(\underline{\alpha})-{m_c \cos\theta_2^\prime t_1\over 16\pi^2}\widetilde{T}_4(\underline{\alpha})\right.
\\ \nonumber &&-\left.{m_c \sin\theta_2^\prime t_2\over 2\pi^2}\widetilde{T}_1(\underline{\alpha})-{m_c \sin\theta_2^\prime t_2\over 2\pi^2}\widetilde{T}_4(\underline{\alpha})+{m_c \cos\theta_2^\prime t_2\over 4\pi^2}\widetilde{T}_1(\underline{\alpha})+{m_c \cos\theta_2^\prime t_2\over 4\pi^2}\widetilde{T}_4(\underline{\alpha})-{7m_c \cos\theta_2^\prime t_1\over 16\pi^2\alpha}\widetilde{T}_2(\underline{\alpha})\right.
\\ \nonumber 
&&-\left.{7m_c \cos\theta_2^\prime t_1\over 16\pi^2\alpha}\widetilde{T}_3(\underline{\alpha})
+{m_c \sin\theta_2^\prime t_2\over 4\pi^2\alpha}\widetilde{T}_1(\underline{\alpha})+{m_c \sin\theta_2^\prime t_2\over 4\pi^2\alpha}\widetilde{T}_4(\underline{\alpha})-{m_c \cos\theta_2^\prime t_2\over 4\pi^2\alpha}\widetilde{T}_1(\underline{\alpha})\right.
\\  &&-\left.{m_c \cos\theta_2^\prime t_2\over 4\pi^2\alpha}\widetilde{T}_4(\underline{\alpha})\Big)\langle\bar s s\rangle\delta(s-{m_c^2\over(1-\alpha)\alpha})\right\}\, .
\\ \rho_2^h(s,M_B)&=&0 \, .
\end{eqnarray}
\end{widetext}
where $e_s$ and $e_c$ are the electric charges of the strange and charm quarks, respectively. The integration limits, parameters $t_1$ and $t_2$, as well as functions $F(\alpha,\beta)$ and $H(\alpha)$ are
\begin{eqnarray}
\nonumber \alpha_{min}&=&{1-\sqrt{1-4 m_c^2\over s}\over 2}\, ,
\\ \nonumber \alpha_{max}&=&{1+\sqrt{1-4 m_c^2\over s}\over 2}\, ,
\\ \nonumber \beta_{min}&=& {\alpha m_c^2\over s\alpha-m_c^2}\, ,
\\ \nonumber \beta_{max}&=& 1-\alpha\, ,
\\ \nonumber 
t_1&=&\sin\theta_2-{\cos\theta_2\over 4}\, ,
\\ \nonumber
t_2&=&-{\cos\theta_2\over8}\, ,
\\ \nonumber \int \mathcal{D}\underline{\alpha}&=&\int_0^1 d\alpha_1\int_0^1 d\alpha_2 \int_0^1 d\alpha_3 \delta(1-\alpha_1-\alpha_2-\alpha_3)\, ,
\\ \nonumber F(\alpha,\beta)&=& m_c^2(\alpha+\beta)-\alpha\beta s\, ,
\\ \nonumber H(\alpha)&=& m_c^2-\alpha(1-\alpha) s\, .
\end{eqnarray}

\section{Transition magnetic moments in the quark model}
\label{app:qm}

In the nonrelativistic constituent quark model, the $3/2^+\to 1/2^+$ radiative transition is dominated by the magnetic-dipole ($M1$) multipole. The transition magnetic moment is defined as
\begin{eqnarray}
	\mu_{fi}&=&\Big\langle B,J_z=\tfrac12\Big|\sum_{i=1}^{3}\mu_i\,\sigma_{iz}\Big|B^*,J_z=\tfrac12\Big\rangle\, ,
	\\ \mu_i&=&\frac{e_i}{2m_i}=\frac{e_i}{e}\,\frac{m_N}{m_i}\,\mu_N\, ,
\end{eqnarray}
where $e_i$ and $m_i$ are the charge and constituent mass of the $i$-th quark, $\sigma_{iz}$ is the Pauli matrix, and $\mu_N=e/(2m_N)$ is the nuclear magneton.

For a doubly heavy baryon $QQq$ with quark ordering $(Q,Q,q)$, the two heavy quarks $Q$ are coupled to spin $S_{QQ}=1$. The spin-flavor wave functions of the initial $3/2^+$ and final $1/2^+$ states with $J_z=1/2$ are
\begin{align}
	&|B^*,J_z=\tfrac12\rangle=\frac{1}{3}(\uparrow \uparrow \downarrow+\uparrow \downarrow \uparrow+\downarrow \uparrow \uparrow)(QQq+QqQ 
\\ \nonumber &~~~~~~~~~~~~~~~~~+qQQ)\, ,
\\	&|B,J_z=\tfrac12\rangle={\sqrt{2}\over3}\left\{{1\over2}(\uparrow \downarrow \uparrow-\downarrow \uparrow \uparrow)(QqQ-qQQ)\right.
\\ \nonumber &~~~~~~~~~~~~~~~~+\left. {1\over2}(\uparrow \uparrow \downarrow-\uparrow \downarrow \uparrow)(QQq-QqQ)\right.
\\ \nonumber &~~~~~~~~~~~~~~~~+\left.{1\over2}(\uparrow \uparrow \downarrow-\downarrow \uparrow \uparrow)(QQq-qQQ)\right\}\, ,
\end{align}
where $Q=c$ or $b$, and $q=u$, $d$, or $s$. Using the constituent quark masses $m_u=m_d=0.336$~GeV, $m_s=0.540$~GeV, $m_c=1.66$~GeV, and $m_b=4.70$~GeV, the resulting transition magnetic moments for the six channels are
\begin{align}
	&\mu(\Xi_{cc}^{*++}\to\Xi_{cc}^{++}\gamma)=\frac{2\sqrt2}{3}(\mu_c-\mu_u)\approx -1.40\,\mu_N\, ,
\\
	&\mu(\Xi_{cc}^{*+}\to\Xi_{cc}^{+}\gamma)=\frac{2\sqrt2}{3}(\mu_c-\mu_d)\approx +1.23\,\mu_N\, ,
\\
	&\mu(\Omega_{cc}^{*+}\to\Omega_{cc}^{+}\gamma)=\frac{2\sqrt2}{3}(\mu_c-\mu_s)\approx +0.90\,\mu_N\, ,
\\
	&\mu(\Xi_{bb}^{*0}\to\Xi_{bb}^{0}\gamma)=\frac{2\sqrt2}{3}(\mu_b-\mu_u)\approx -1.82\,\mu_N\, ,
\\
	&\mu(\Xi_{bb}^{*-}\to\Xi_{bb}^{-}\gamma)=\frac{2\sqrt2}{3}(\mu_b-\mu_d)\approx +0.81\,\mu_N\, ,
\\
	&\mu(\Omega_{bb}^{*-}\to\Omega_{bb}^{-}\gamma)=\frac{2\sqrt2}{3}(\mu_b-\mu_s)\approx +0.48\,\mu_N\, .
\end{align}

\bibliographystyle{elsarticle-num}
\bibliography{ref}

\end{document}